\documentclass{article}

\usepackage[preprint]{neurips_2025}

\usepackage[utf8]{inputenc} % allow utf-8 input
\usepackage[T1]{fontenc}    % use 8-bit T1 fonts
\usepackage[pagebackref,breaklinks,colorlinks]{hyperref}
\usepackage{url}            % simple URL typesetting
\usepackage{booktabs}       % professional-quality tables
\usepackage{amsfonts}       % blackboard math symbols
\usepackage{nicefrac}       % compact symbols for 1/2, etc.
\usepackage{microtype}      % microtypography
\usepackage{xcolor}         % colors
\usepackage{amsmath}
\usepackage{amssymb}
\usepackage{graphicx}
\usepackage{booktabs}
\usepackage[normalem]{ulem}
\usepackage{multirow}
\usepackage{subcaption}
\usepackage{enumitem}
\usepackage{cleveref}
\usepackage{minitoc}

\definecolor{fab-green}{rgb}{0.192, 0.639, 0.329}
\definecolor{core-orange}{rgb}{0.90, 0.33, 0.05}

\newcommand{\methodName}{CryoSPIRE}
\newcommand{\zSpace}{{\mathfrak Z}}
\newcommand{\fSpace}{{\mathfrak F}}

\title{Reconstructing Heterogeneous Biomolecules via Hierarchical Gaussian Mixtures and Part Discovery}

\author{
    Shayan Shekarforoush\normalfont{\textsuperscript{1, 2}} \\ \texttt{shayan@cs.toronto.edu} \\ \And
    David B. Lindell\normalfont{\textsuperscript{1, 2}} \\ \texttt{lindell@cs.toronto.edu} \\ 
    \AND
    Marcus A. Brubaker\normalfont{\textsuperscript{1, 2, 3}} \\ \texttt{mab@eecs.yorku.ca} \\ \And
    David J. Fleet\normalfont{\textsuperscript{1, 2}} \\ \texttt{fleet@cs.toronto.edu} \\ \And
    \normalfont{
    \textsuperscript{1}University of Toronto 
    \quad 
    \textsuperscript{2}Vector Institute 
    \quad 
    \textsuperscript{3}York University
    }
}

\begin{document}

\maketitle

\begin{abstract}
\vspace*{-0.35cm}

Cryo-EM is a transformational paradigm in molecular biology where computational methods are used to infer 3D molecular structure at atomic resolution from extremely noisy 2D electron microscope images.
At the forefront of research is how to model the structure when the imaged particles exhibit non-rigid conformational flexibility and compositional variation where parts are sometimes missing.
We introduce a novel 3D reconstruction framework with a hierarchical Gaussian mixture model, inspired in part by Gaussian Splatting for 4D scene reconstruction. 
In particular, the structure of the model is grounded in an initial process that infers a part-based segmentation of the particle, providing essential inductive bias in order to handle both conformational and compositional variability.  The framework, called \methodName, is shown to reveal biologically meaningful structures on complex experimental datasets, and establishes a new state-of-the-art on CryoBench, a benchmark for cryo-EM heterogeneity methods.
\href{https://shekshaa.github.io/CryoSPIRE}{Project Webpage}.

\end{abstract}
\vspace*{-0.2cm}
\section{Introduction}
\vspace*{-0.25cm}

\begin{figure}[t]
    \centering
    \includegraphics[width=0.97\linewidth]{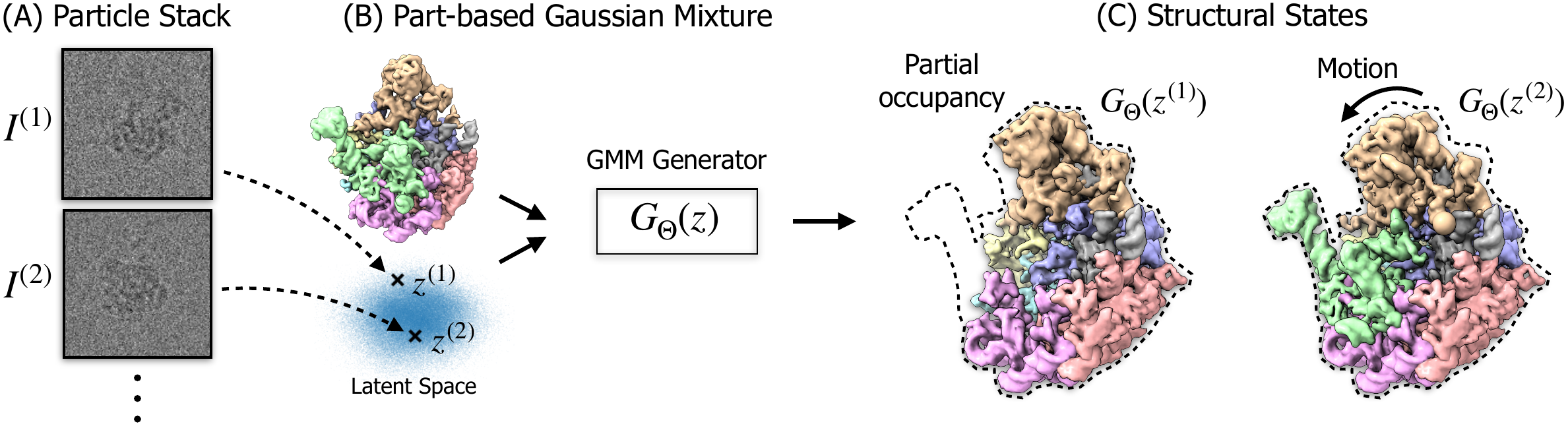}
   \vspace*{-0.1cm}
   \caption{
    (A) Based on a stack of noisy particle images, (B) \methodName{} learns a part-based Gaussian mixture, with parameters $\Theta$, 
    % each part of which is  Gaussian mixture, 
    and a latent space representing structural heterogeneity. 
    Given a latent code $z$, a generator produces a 3D density map.
    (C) The model supports compositional variability (e.g., $G_{\Theta}(z^{(1)})$ with a missing part), and conformational flexibility (e.g., $G_{\Theta}(z^{(2)})$ with part deformation).
    }
    \vspace*{-0.2cm}
    \label{fig:teaser}
\end{figure}

Single-particle cryo-electron microscopy (cryo-EM) is a computationally driven experimental paradigm that is transforming molecular biology by enabling 3D structure determination of biomolecules, such as proteins and viruses, at near-atomic resolutions~\cite{cryoSPARC-CVPR2015,ResolutionRevolution,SCHERES2016125}.
The core computational task is estimating a 3D structure from 2D images with unknown orientation and position, under extremely low signal-to-noise conditions.
Essential to their biological function, biomolecules exhibit varying degrees of {\em conformational flexibility}, where structures deform non-rigidly, and {\em compositional variation}, where parts of a structure may be present in some images and absent in others (see Fig.~\ref{fig:teaser}).
Accordingly, a major challenge in cryo-EM is the estimation of 3D structures from such heterogeneous data and, to that end, how to infer meaningful representations of structures such as parts that capture their heterogeneity.
% variability.
The crux of this challenge is how to effectively represent and regularize this variability without overfitting to the noise in cryo-EM images.
Existing methods, while encouraging, are generally limited in either expressiveness, interpretability, or efficiency.

Here, we propose \methodName{}, a new method for heterogeneous reconstruction.
We leverage a part-based Gaussian mixture model (GMM) of 3D density that enables \methodName{} to represent both conformational and compositional heterogeneity, unlike some existing deformation-based methods~\cite{herreros2023estimating,3DFlex-NM-2023}.
Further, it provides a naturally interpretable and physically plausible, part-based structure in contrast to existing latent variable methods based on linear density subspaces~\cite{Recovar-2025,3DVA2021} or neural field models~\cite{levy2024revealing,levy2022amortized,cryoDRGN2021}.
A key challenge with part-based GMMs  concerns initialization and the discovery of parts.
We propose a novel method for part discovery which estimates a coarse-grained GMM with per-Gaussian learnable features (c.f., \cite{bae2024per}) and an MLP which defines Gaussian locations and amplitudes.
We show that these learned features naturally encode characteristics of structural heterogeneity, which we leverage to infer a part-based segmentation of the structure.
Inspired in part by Scaffold-GS~\cite{lu2024scaffold}, we define \methodName{} (Scaffold Part-Aware Mixture of Gaussians), a hierarchical model which estimates a Gaussian mixture wherein the composition of components and their deformation are defined in terms of a set of anchors, corresponding to parts.
The resulting model naturally allows for the arbitrary combination of parts which can both rigidly move and locally deform as a function of an input heterogeneity latent code (see Fig.\ \ref{fig:teaser}).
To our knowledge, this is first GMM-based model to be successfully benchmarked on 
% We are first to benchmark a GMM-based model on 
CryoBench~\cite{jeon2024cryobench}, a standardized benchmark for cryo-EM heterogeneity with ground-truth labels.
In particular, \methodName{} outperforms widely used and state-of-the-art methods \cite{Recovar-2025,levy2024revealing,3DVA2021,3DFlex-NM-2023,cryoDRGN2021}, sometimes by a wide margin.
% On CryoBench~\cite{jeon2024cryobench}, a standardized benchmark for cryo-EM heterogeneity with ground-truth labels, \methodName\ outperforms widely used and state-of-the-art methods \cite{Recovar-2025,levy2024revealing,3DVA2021,3DFlex-NM-2023,cryoDRGN2021}, sometimes by a wide margin.
% By contrast, previous GMM-based heterogeneity methods have not been successfully benchmarked.
Through ablations, we also validate key design choices, demonstrating the benefits of Gaussian features over positional encoding as in DynaMight~\cite{DynaMight-2024}, and highlighting the benefits of hierarchical motion modeling.
Finally, on real experimental cryo-EM data, \methodName{} automatically discovers representations of 3D density maps that correspond to   biologically meaningful parts.

To summarize our contributions:
% We propose a new part-discovery method for unsupervised segmentation of 3D biomolecular structures based on a coarse-grained GMM.
we propose a new method enabling part-discovery on 3D biomolecular structures based on a coarse-grained GMM.
This part-based structure is used to initialize a novel, hierarchical GMM-based model for heterogeneous reconstruction with compositional and conformational variability.
The resulting framework, \methodName{}, establishes a new state-of-the-art on quantitative benchmarks and qualitative experimental datasets.

\vspace*{-0.2cm}
\section{Background and Related Work}
\vspace*{-0.25cm}
\label{sec:related-work}

{\bf Latent Variable Models.}
Heterogeneous cryo-EM reconstruction methods typically introduce latent variables to represent structural variability of the 3D density map.
3DVA~\cite{3DVA2021} and RECOVAR~\cite{Recovar-2025} learn a linear subspace to represent variation in 3D density maps, with clever numerical and regularization techniques to optimize high-dimensional basis maps at high spatial resolutions.
Such methods are often notoriously memory intensive and have limited expressiveness due to the linear subspace. 
Much current work has shifted to nonlinear latent models and deep learning \cite{CryoFeren2023,levy2024revealing,cryoDRGN2021}, with Cryo-DRGN~\cite{cryoDRGN2021} and DRGN-AI~\cite{levy2024revealing} using auto-encoders to obtain latent codes and conditional coordinate networks~\cite{MildenhallEtAl2022} to generate density maps.
Such latent-variable models are hard to interpret, however, as conformational and compositional heterogeneity are not decoupled, and they provide no explicit model of motion between conformational states.
By contrast, the latents in 3DFlex~\cite{3DFlex-NM-2023} encode flow fields that model the conformational deformation of a canonical structure.
While resolving detailed motion and improving the quality of density maps, 3DFlex cannot handle compositional heterogeneity, and it
% is hard to regularize, 
is highly sensitive to regularization,
often requiring substantial trial and error.

{\bf GMM-Based Methods.}
Gaussian mixtures have been used to model 3D density~\cite{Chen2021GMM,MuyuanChen-GMM-2023,MuyuanChen-GMM-2023b,DynaMight-2024}; they provide a sparse, compact representation in which conformation and compositional variability are modeled in terms of positions and amplitudes of Gaussian components.
With Gaussian components viewed as atomic primitives, such models also facilitate physics-based priors~\cite{MuyuanChen-GMM-2023b,DynaMight-2024} and subsequent molecular model fitting.
Nevertheless, existing GMM-based methods fall short in various ways.
E2GMM~\cite{Chen2021GMM} and related methods~\cite{MuyuanChen-GMM-2023,MuyuanChen-GMM-2023b} generate GMM parameters with a single network, which scales poorly to large numbers of Gaussians.
Further, their multi-scale smoothness priors~\cite{MuyuanChen-GMM-2023b} are based on an arbitrary hierarchy which fails to capture part-based structures, thus resorting to manual part masks to resolve and estimate local motions. 
% as \methodName{} does.
% To isolate localized motions, \cite{MuyuanChen-GMM-2023b} resorts to manual masking while our method enables automatic identification of coherent regions.
% Regularization through smoothness priors 
% Although regularizing the motion at multiple scales, \cite{MuyuanChen-GMM-2023b} defines an arbitrary hierarchy which does not conform to the underlying part organization, hence manual part masks are required.
DynaMight~\cite{DynaMight-2024} is similar to \methodName{} in defining an explicit motion model, but it is unable to handle compositional variations, and, as we show, its positional encodings are inferior to our learnable features.

% Alternatively, Gaussian Mixtures has emerged as a 3D density model~\cite{Chen2021GMM,MuyuanChen-GMM-2023,MuyuanChen-GMM-2023b,DynaMight-2024}, offering several nice properties:
% i) thanks to the discrete nature, they sparsely represent the density map unlike the volumetric representations,
% ii) they support explicit representations of motion of Gaussian positions and compositional occupancy through variations in Gaussian amplitudes. 
% iii) they facilitate physics-based priors~\cite{MuyuanChen-GMM-2023b,DynaMight-2024} and subsequent molecular model fitting.

% Nevertheless, GMM-based models to date fall short in different ways.
% E2GMM~\cite{Chen2021GMM} and follow-up work~\cite{MuyuanChen-GMM-2023,MuyuanChen-GMM-2023b} are not scalable to a larger number of components, as parameters of all Gaussians are generated as a single high-dimensional output by a network.
% Although regularizing the motion at multiple scales, \cite{MuyuanChen-GMM-2023b} defines an arbitrary hierarchy which does not conform to the underlying part organization.
% To isolate localized motions, \cite{MuyuanChen-GMM-2023b} uses manual masking while our method enables automatic identification of coherent regions.
% Finally, our method is similar to DynaMight~\cite{DynaMight-2024} in defining an explicit motion model.
% However, DynaMight does not capture compositional heterogeneity and uses Gaussian positions to condition networks, whereas we adopt learnable features that arguably store less spatially biased and more relevant heterogeneity information for individual Gaussians.

{\bf Gaussian Splatting.}
Beyond cryo-EM, the effectiveness of GMMs has been demonstrated in 3D Gaussian Splatting~\cite{Kerbl2023tdgs,zwicker2002ewa}, a technique which provides a fast approximation to the volume rendering integral~\cite{drebin1988volume,max2002optical}, enabling efficient high-fidelity reconstruction of 3D scenes from multi-view images~\cite{guedon2024sugar,kheradmand20243d,luiten2024dynamic,wu20244d,yang2024deformable,yu2024mip}.
3D Gaussian Splatting represents scene appearance and structure using thousands to millions of Gaussian components, each associated with parameters that control opacity and view-dependent color.
\methodName{} is in part inspired by Gaussian Splatting~\cite{bae2024per,lu2024scaffold}, but tailored to cryo-EM, with a different image formation model, images with signal-to-noise ratios less than 5\%, and a novel method for part discovery.

{\bf GMM Image Formation.}
Following \cite{Chen2021GMM,MuyuanChen-GMM-2023,MuyuanChen-GMM-2023b,DynaMight-2024}, we parameterize the terms of a Gaussian mixture with center $\boldsymbol{c} \in \mathbb{R}^3$, isotropic scale $s \in \mathbb{R}$, and an amplitude $m \in \mathbb{R}$:
% The 3D density at location $\boldsymbol{p} \in \mathbb{R}^3$ is given by
\begin{equation}
    f(\boldsymbol{p}) ~=~ \sum_{i} m_i \exp\left(-\frac{||\boldsymbol{p} - \boldsymbol{c}_i||^2_2}{2s_i^2}\right) \ ,
    \label{eq:density-map}
\end{equation}
for location $\boldsymbol{p} \in \mathbb{R}^3$.
We transform the GMM into the observation space 
%(microscope coordinates) 
for the $n$-th particle image, with a rotation $\boldsymbol{R}^{(n)} \in SO(3)$ and translation $\boldsymbol{t}^{(n)}  \in \mathbb{R}^3$,
% =(t_x,t_y,t_z)$, 
% as $\hat{\mathcal{G}}^{(n)} =\{(R p_i + t, s_i, a_i)\}$. 
followed by an integral projection along the $z$-axis of the microscope, to obtain a noise-free 2D image, $\tilde{I}(\tilde{\boldsymbol{p}})$, \cite{Chen2021GMM}:
\begin{equation}
    \tilde{I}^{(n)}(\tilde{\boldsymbol{p}}) ~=~ \sum_{i} \sqrt{2\pi} s_i m_i \exp\left(-\frac{|| \,\tilde{\boldsymbol{p}} - [\boldsymbol{R}^{(n)} \, \boldsymbol{c}_i + \boldsymbol{t}^{(n)} ]_{xy} \, ||^2_2}{2s_i^2}\right) \ 
    \label{eq:render} \ ,
\end{equation}
where $\tilde{\boldsymbol{p}} \in \mathbb{R}^{2}$ and
$[\cdot]_{xy}$ is an operator to discard $z$ coordinate of the input position.
Cryo-EM images are then convolved with microscope point spread function and corrupted by additive mean-zero Gaussian noise,
$\hat{I}^{(n)} = g^{(n)} \star \tilde{I}^{(n)} + \epsilon^{(n)}$.
Like other cryo-EM models, the parameters are typically optimized by minimizing a squared L2 reconstruction loss between model predictions and observed images.
See the supplement for more details on image formation and the image likelihood.
\vspace*{-0.2cm}
\section{\methodName{}}
\vspace*{-0.25cm}

% \DavidF{Somewhere we need to explain that there is a latent space and, conditioned on a latent code, a 3D density generater.
% This could be above in background.  Then here we just focus on the density model, conditioned on latent code.  What follows below doesn't introduce this broader context. We also need to say that we assume a particle stack with per-particle poses.}
% \Shayan{I am introducing that in the second paragraph. I thought first we need to motivate what we want to achieve in this problem and then describe the problem setup.}
% \DavidF{Agreed.  But the latent heterogeneous landscape is critical to the problem, as are the initial poses.  In this section we are not describing the larger problem but rather just the conditional GMM density map model.  But that presupposes a lot of context.}
% \Shayan{I agree. Those other contexts are not sufficiently described here. Not sure if background is the right place to explain in more details or not}
% \DavidF{Let's make sure that the general framework, including a parameterized density map, a latent space, and a generator that produces a density given a latemt code, etc are included in the teaser and background since they are common to all other methods.}
% \Shayan{Sure, I will include all these components in the teaser figure and refer to them in the text.}
% \DavidF{SG}

Heterogeneous cryo-EM involves non-rigid 3D reconstruction from noisy 2D images. 
For such an inverse problem, regularization and inductive bias are key.  Local smoothness is a natural choice for regularization, however, smoothness alone is not sufficient as nearby regions can deform in somewhat independent ways~\cite{3DFlex-NM-2023}. 
Further, the presence or absence of biomolecule parts is not dictated by spatial proximity alone.
Macromolecular complexes, like many objects, naturally possess a part-based structure that connects to their compositional and conformational variations.
But a coherent 3D part-decomposition is unavailable \emph{a priori}, and estimating parts from noisy 2D observations is inherently challenging. 
As a consequence, prior work resort to manually designed masks or meshes~\cite{nakane2020multi,3DFlex-NM-2023}.

% Here, we assume that we are given a particle stack of images, a per-particle pose estimate, and an initial, albeit crude, 3D structure.
% The goal is to learn a density model that specifies a structural state conditioned on a latent code that lies in a space embedding the heterogeneity landscape.
% The goal of learning is to learn a latent space that 
% captures the landscape of heterogeneity, a per-particle latent code $z^{(n)}$, and a density model that specifies a 3D density map conditioned on the latent code. \DavidF{point to teaser here?}
% In this paper 
% We focus on the nature of the GMM-based 3D density model, and how heterogeneity is parameterized within the model.

\begin{figure}[t]
    \centering
    \includegraphics[width=\linewidth]{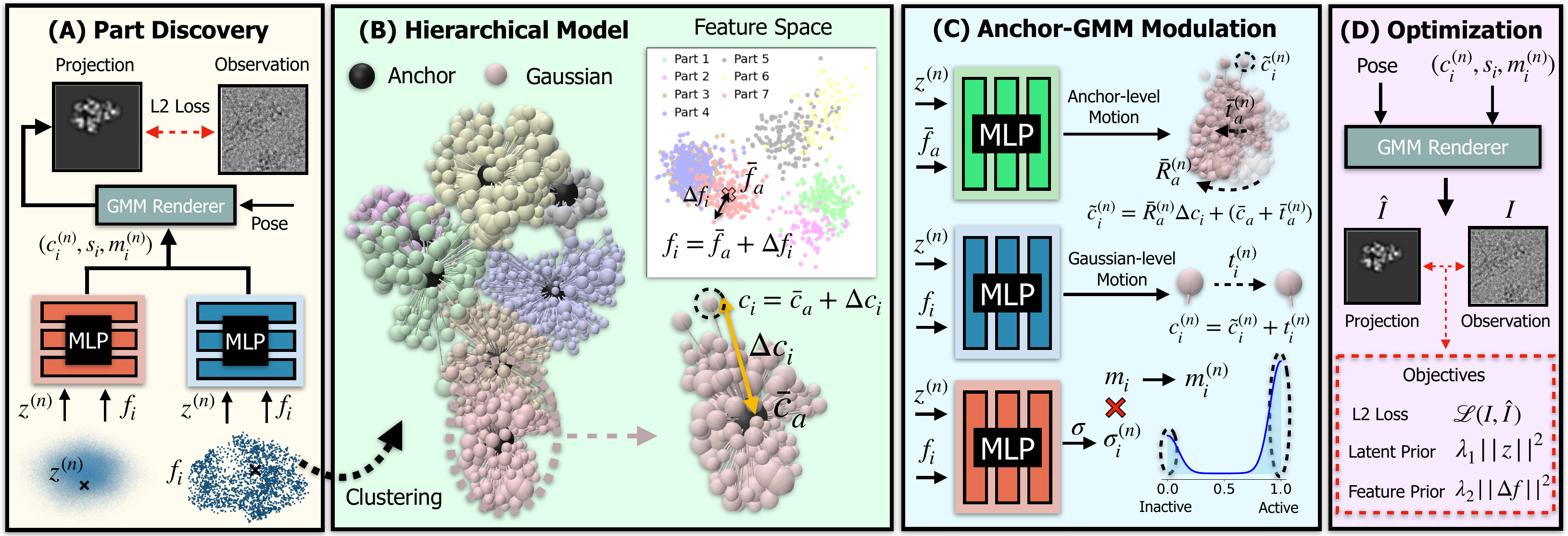}
    \caption{
    Overview of \methodName.
    \textbf{(A)} To infer parts, we optimize a coarse GMM with neural networks that generate Gaussian amplitudes and centers, conditioned on image latent codes and Gaussian features.
    \textbf{(B)} Clustering on learned Gaussian features provides meaningful parts. 
    The \methodName{} model comprises one anchor and a set of Gaussians per part.
    \textbf{(C)} MLPs generate the rigid-body motion of each anchor (top), per-Gaussian displacements relative to the anchor frames (middle), and per-Gaussian activations in (0,1) to represent occupancy (bottom).
    \textbf{(D)} A reconstruction loss compares observed images to 2D projection of the corresponding 3D GMM.
    Priors encourage bounded latent code and small feature offsets.
    %  Overview of our framework.
    % \textbf{(A)} To infer parts, we optimize a coarse-grained model with standard L2 loss, and obtain two neural networks that modulate the amplitude and the centers of Gaussians, conditioned on image latent codes and Gaussian features.
    % \textbf{(B)}
    % The learned Gaussian features form groups corresponding to semantically meaningful parts, which are identified by clustering.
    % Accordingly, we build the hierarchical model consisting of anchors and Gaussian components. 
    % Each anchor is defined based on the cluster center, while Gaussian features and centers are parameterized relative to the anchors. 
    % \textbf{(C)}
    % Two MLPs, respectively, estimate the anchor-level motion, parameterized by rigid-body transformations (top), and Gaussian-level motion with per-Gaussian deformations (middle).
    % A separate MLP is also used to estimate active and inactive Gaussians to account for compositional occupancy (bottom).
    % \textbf{(D)}
    % Given the modulated GMM and the particle pose, we obtain a projection which is penalized against the observation using L2 reconstruction loss.
    % Also, priors encourage bounded latent code and small feature offsets.
    }
    \vspace*{-0.1cm}
    \label{fig:method}
\end{figure}

Here, we propose a novel two-stage GMM-based framework.
Given particle images with corresponding poses $\{ (I^{(n)}, \boldsymbol{R}^{(n)}, \boldsymbol{t}^{(n)}) \}_{n=1}^N$, and a crude initial 3D structure, we first optimize a coarse-grained GMM in which each Gaussian component is augmented with a learnable feature vector (c.f., \cite{bae2024per}).
We observe that the learned features encode meaningful information about structural regularities.
In particular, Gaussian components that coherently deform or consistently appear or disappear receive similar features, facilitating the inference of a part-based segmentation of the particle. 
Second, based on the identified parts and inspired by Scaffold-GS~\cite{lu2024scaffold}, we define a part-aware Gaussian mixture model in terms of a set of anchors, one per part, each with a corresponding set of Gaussians.
% Second, based on the identified parts, and inspired by Scaffold-GS~\cite{lu2024scaffold}, we define our final model, Scaffold Part-Aware Mixture of Gaussians (\methodName), 
% which consists of constellations of Gaussian components corresponding to parts, each represented by an anchor.
Optimizing this representation recovers a high-resolution representation of 3D density maps with compositional and conformational variability.
% We embed anchors in the same feature space and tie each Gaussian in parameterization to the corresponding anchor.
% The dense sets of Gaussian components, conditioned on their corresponding anchors, then provide fine-grained structural fidelity.
In what follows, we describe the part-based hierarchical model, (Fig.~\ref{fig:method}B--D), followed by the part discovery method and initialization scheme (Fig.~\ref{fig:method}A).

\vspace*{-0.25cm}
\subsection{Part-Aware Gaussian Mixture} 
\vspace*{-0.2cm}

Our hierarchical model is conditioned on a latent coordinate $\boldsymbol{z} \in \zSpace \subset \mathbb{R}^D$ for each image, which specifies the state of the macromolecule.
The density model itself comprises a set of anchors, each associated with a meaningful part of the  macromolecule (Fig.~\ref{fig:method}B).
We parameterize the anchors as, ${\cal A} ~=~ \{(\bar{\boldsymbol{c}}_a, \bar{\boldsymbol{f}}_a)\}_{a=1}^{A}$,
% \begin{equation}
    % {\cal A} ~=~ \{(P_j, F_j)\}_{j=1}^{K} \ ,
    % \label{eq:anchors}
% \end{equation}
where $\bar{\boldsymbol{c}}_a \in \mathbb{R}^3$ specifies the anchor center location in a canonical frame, and $\bar{\boldsymbol{f}}_a \in \fSpace \subset \mathbb{R}^{E}$ is an associated feature vector that encodes heterogeneity information of its corresponding part.
The GMM has $G$ Gaussian components associated with anchors (Fig.~\ref{fig:method}B, left), denoted by ${\cal G} = \{(\boldsymbol{f}_i, \boldsymbol{c}_i, s_i, m_i, a_i )\}_{i=1}^{G}$ where $\boldsymbol{f}_i \in \fSpace$ and $ a_i \in \{1, \dots , A\} $ specifies the anchor associated with the Gaussian that is set by the part discovery method below.
% \begin{equation}
%     {\cal G} = \{(p_i, s_i, a_i, f_i)\}_{i=1}^{M} \ .
%     \label{eq:coarse-gmm}
% \end{equation}
% Here, the conventional GMM is augmented with learnable features, $\boldsymbol{f}_i \in \mathbb{R}^{E}$.
% Each Gaussian is associated with an anchor denoted by $ a_i \in \{1, \dots , A\} $ which is set by the part discovery method described below.
We parameterize the position and feature embedding of the $i$-th Gaussian relative to its associated anchor $a_i$ as
\begin{equation}
    \boldsymbol{c}_i = \bar{\boldsymbol{c}}_{a_i} + \Delta \boldsymbol{c}_i \ , \quad \boldsymbol{f}_i = \bar{\boldsymbol{f}}_{a_i} + \Delta \boldsymbol{f}_i \ ,
    \label{eq:anchor-gaussian-binding}
\end{equation}
where $\Delta \boldsymbol{c}_i \in \mathbb{R}^3$ and $\Delta \boldsymbol{f}_i \in \mathbb{R}^{E}$ are learnable offsets.
We initially set $\Delta \boldsymbol{f}_i=\boldsymbol{0}$ so all Gaussians are initialized with the features of their corresponding anchors. 

% \paragraph{Modeling Heterogeneity.}
% Given $({\cal A}, {\cal G})$ as a shared hierarchical representation, we modulate it to capture both conformational and compositional variability in various particle images (Fig.~\ref{fig:method}C).
To enable conformational variability, we parameterize deformations at two levels.
First, the large-scale motion of each anchor frame is parameterized as a rigid body transformation (Fig.~\ref{fig:method}C, top).
Given the latent code for $n$-th particle image, $\boldsymbol{z}^{(n)} \in \zSpace$, and the anchor feature vector $\bar{\boldsymbol{f}}_{a_i}$, we compute the rotated and translated center of the $i$-th Gaussian, $\tilde{\boldsymbol{c}}^{(n)}_i$, as
\begin{equation}
    \tilde{\boldsymbol{c}}^{(n)}_i = \bar{\boldsymbol{R}}^{(n)}_{a_i} \Delta \boldsymbol{c}_i + (\bar{\boldsymbol{c}}_{a_i} + \bar{\boldsymbol{t}}^{(n)}_{a_i}) \ , \quad \mbox{where} ~~ \bar{\boldsymbol{R}}^{(n)}_{a_i}, \bar{\boldsymbol{t}}^{(n)}_{a_i} = \text{MLP}^{\cal A}([\bar{\boldsymbol{f}}_{a_i},\boldsymbol{z}^{(n)}]; W^{\cal A}) \ ,
    \label{eq:anchor-motion}
\end{equation}
where $[\cdot, \cdot]$ denotes concatenation, and the MLP with weights $W^{\cal A}$ returns a rotation $\boldsymbol{R}^{(n)}_{a_i} \in SO(3)$ and translation vector $\boldsymbol{t}^{(n)}_{a_i} \in \mathbb{R}^{3}$.
% , which specify the orientation of Gaussians relative to the anchor and the global position of the anchor.
To capture fine-scale flexibility, additional shifts are applied to individual Gaussians (Fig.~\ref{fig:method}C, middle), i.e.,
\begin{equation}
    \boldsymbol{c}^{(n)}_i = \boldsymbol{\tilde{c}}^{(n)}_i +  \boldsymbol{t}^{(n)}_i \ , \quad \mbox{where} ~~ \boldsymbol{t}^{(n)}_{i} = \text{MLP}^{\cal G}_{\boldsymbol c}([\boldsymbol{f}_{i}, \boldsymbol{z}^{(n)}]; W^{\cal G}_{\boldsymbol c}) \ .
    \label{eq:gaussian-motion}
\end{equation}
Here, the network $\text{MLP}^{\cal G}_{\boldsymbol c}$, with separate weights $W^{\cal G}_{\boldsymbol c}$, generates individual Gaussian displacements, $\boldsymbol{t}^{(n)}_i \in \mathbb{R}^{3}$, which are smooth as Gaussians associated with the same anchor will have similar features.

Finally, to account for compositional variability, where regions of a density map may be missing, we modulate Gaussian amplitudes (Fig.~\ref{fig:method}C, bottom), as
\begin{align}
    m^{(n)}_i &= m_i \times \sigma^{(n)}_i \ ,  
    \quad
    \mbox{where} ~ \sigma^{(n)}_i = \text{MLP}^{\cal G}_m([\boldsymbol{f}_i,\boldsymbol{z}^{(n)}]; W^{\cal G}_{m}) \ .
    \label{eq:amp-modulation}
\end{align}
Here, $\text{MLP}^{\cal G}_m$ is an MLP with a sigmoid output activation to restrict the modulation to $(0,1)$.
Values close to $0$ and $1$, respectively, correspond to inactive (absent) and active (present) Gaussians.
Considering both modifications to centers and amplitudes, we obtain a modulated set of 3D Gaussians for $n$-th particle image, $ \mathcal{G}^{(n)} = \{(\boldsymbol{c}^{(n)}_i, s_i, m^{(n)}_i)\}$.
Gaussian scales remain the same as they control local resolution, a factor independent of structural variability.

% \paragraph{Optimization.} 
% \Marcus{Shayan, don't we also optimize wrt $\delta c_i$, $\bar{c}_a$, etc?}
% Next, we compute the integral projection, $\hat{I}^{(n)}$, using Eq.~\ref{eq:render}, and
We jointly optimize the parameters $\Theta$ (which includes Gaussian and anchor parameters and MLP weights),
% $$=\{\{\bar{f}_a\}_a, \{\Delta f_i , s_i, m_i, \Delta c_i\}_{i}, W^{\cal G}_{\boldsymbol{c}}, W^{\cal G}_{m}, W^{\cal A}\}$, 
and the per-image latent coordinates, $Z=\{z^{(n)}\}$,  by minimizing the objective (Fig.~\ref{fig:method}D)
%
% \DavidF{Define parameters $\theta$ and express loss as function of $\theta$ and $z$.}
\begin{equation}
    L(\Theta, Z) ~=~ \frac{1}{N}\sum_{n=1}^{N}\mathcal{L}\left(I^{(n)},\hat{I}^{(n)}\right) \, + \, \lambda_z  \, ||\boldsymbol{z}^{(n)}||^2_2 \, +\,  \lambda_f \sum_{i=1}^{G}||\Delta \boldsymbol{f}_i||_2^2 ~ \ ,
    \label{eq:loss}
\end{equation}
where the reconstruction loss, $\mathcal{L}$, is proportional to the negative image log-likelihood (i.e., the squared error between $I^{(n)}$ and $g^{(n)} \star \hat{I}^{(n)}$ where $g^{(n)}$ is the microscope point spread function and $\hat{I}^{(n)}$ is the 2D projection of $\mathcal{G}^{(n)}$ from Eq.~\ref{eq:render}).
% and the image latent codes are updated in auto-decoding fashion~\cite{park2019deepsdf}.
The second term imposes a zero-mean Gaussian prior over the per-image latent codes, ensuring latent coordinates remain bounded~\cite{park2019deepsdf,3DFlex-NM-2023}, while the third term regularizes Gaussians to remain close to the anchor in the feature space.
$\lambda_z$ and $\lambda_f$ are hyperparameters that  control the relative strength of these priors.

\vspace*{-0.25cm}
\subsection{Part Discovery for Model Initialization}
\vspace*{-0.25cm} 

The part discovery process is illustrated in Fig.~\ref{fig:method}A. 
We optimize a coarse-grained model without anchors and with fewer Gaussians, similarly parameterized as ${\cal G} = \{(\boldsymbol{f}_i, \boldsymbol{c}_i, s_i, m_i )\}_{i=1}^{G}$.
Here, the Gaussian features, $\boldsymbol{f}_i$, are directly learnable parameters (and randomly initialized).
We use $\text{MLP}^{\cal G}_c$ (Eq.~\ref{eq:gaussian-motion}), to shift Gaussian centers and $\text{MLP}^{\cal G}_m$ (Eq.~\ref{eq:amp-modulation}) to modulate Gaussian amplitudes.
The parameters are estimated using the L2 reconstruction loss and the latent prior, similar to the objective in Eq.~\ref{eq:loss}.
Once optimized, we find that the feature space naturally groups Gaussians into 3D parts that undergo consistent motion or appear and disappear together.
Remarkably, this property emerges without any direct supervision on features.

To obtain parts, we apply clustering on the Gaussian features, thereby finding regions with reasonably consistent motion and presence.
We then further divide these clusters by clustering in 3D space to ensure large parts are well-covered with anchors.
For clustering we simply use k-means++~\cite{kmeans2007}.
We use the position and feature vector of the Gaussian closest to the centroid of the cluster to initialize the anchor set, ${\cal A} = \{(\bar{\boldsymbol{c}}_a, \bar{\boldsymbol{f}}_a)\}_{a=1}^{A}$.
From the coarse-grained model, we also compute an improved density map which is used to seed the Gaussians of the part-aware model.
This provides a more robust initialization, especially in the presence of large-scale motion which can lead to blurred or over-dispersed density.
Lastly, the coarse-grained model provides a preliminary estimate of the image latent codes, which are used to initialize latent codes in the part-aware model.

{\bf Remark.} Methods for 4D scene reconstruction~\cite{park2021nerfies,pumarola2021d}, and 
DynaMight~\cite{DynaMight-2024} in cryo-EM, often use neural networks to output deformations or motion.
However, they condition on positional encodings of input coordinates instead of learnable features.
Such fixed conditioning strongly biases deformations to be spatially smooth, whereas our approach with learnable feature space enables a more flexible form of piecewise smoothness, allowing nearby parts to move quite differently.
% , which makes it challenging to distinguish a moving part from its nearby rigid parts.
% In contrast, our feature embeddings can adapt to arrange Gaussians based on the underlying semantic parts. 
Through an ablation study, we show that positional encodings quantitatively underperform as well.
% leads to inferior performance.
% \Marcus{I think it's more than this.  Making the deformation conditional on a positional embedding strongly biases the deformation towards spatially coherent deformations.  This makes it hard for there to be moving parts connected to (or close to) rigid ones.}
% \Shayan{@Marcus Does above address your point?}

% \Marcus{The initialization we use is more crude than I realized.  What resolution map are we using?  I wonder if it might work using random back projection as the consensus structure, or even the result of a standard, single class ab-initio run which would turn this part into an entirely ab-initio method.  Would be interesting to see the thresholded consensus densities as well.}
% \DavidF{Agreed. The less reliant on a homogeneous reconstruction the better IMO .... ie we prefer to be closer to ab initio.}
% \Shayan{I guess seeding Gaussians based on the ab-initio should be fine in general. However, we lose the accuracy in poses which significantly affects the final performance. This is shown in CryoBench where ab-intio methods are performing way worse.}

\vspace*{-0.2cm}
\section{Experimental Setup}
\vspace*{-0.25cm}

% \DavidF{We'll need a discussion somewhere on performance measures, including FSC vs GT at 0.5, and half-map FSC at 0.143.  We could split this and have the FSC vs GT in the CryoBench section, with a brief discussion of how the synth data was created.  We can put details in an appendix.}
%Shayan: I think this is addressed

% \Shayan{CryoBench has introduced five synthetic datasets. We are benchmarking on three of them. Among that other two, one is Tomotwin which is a mixture of 100 completely different structures that have nothing in common. So there is no consensus map that we can build on. GMM models and 3DFlex cannot naturally address that.
% Another data is Spike-MD. However, GT data is not available to compute FSC metric. People keep asking in the github issues for GT but no response yet how to generate them.
% Should we be clear up front why we could not evaluate on that two datasets? Or that could raise an unnecessary concern?}
% \DavidF{Let's say we compare against baselines on the cryoBench datasets with  continuous conformational and compositional heterogeneity with GT for quantitative eval.  I think an explanation for why we leave two out is likely unnecessary, or if we do include it, let's put it in the appendix.}
We quantitatively compare \methodName{} with the state-of-the-art methods, namely, RECOVAR~\cite{Recovar-2025}, CryoDRGN~\cite{cryoDRGN2021}, DRGN-AI~\cite{levy2024revealing}, 3DFlex~\cite{3DFlex-NM-2023} and 3DVA~\cite{3DVA2021} using the CryoBench benchmark~\cite{jeon2024cryobench}.
We also provide qualitative results on experimental datasets.
% and ablate some methodological decisions.

{\bf CryoBench.} 
The sole benchmark for cryo-EM heterogeneity is CryoBench~\cite{jeon2024cryobench}, a set of synthetic datasets with ground-truth labels and a protocol for quantitative evaluation.
Two datasets, 
% From Cryo use the IgG-1D, IgG-RL and Ribosembly synthetic datasets from CryoBench~\cite{jeon2024cryobench}, 
{\em IgG-1D} and {\em IgG-RL}, are based on the human immunoglobulin G (IgG) complex, simulating conformational changes by rotating the dihedral angle between the Fab domain and the IgG core (see Fig.~\ref{fig:IgG-1D}D), generating $100$ distinct conformations, each with $1{,}000$ particle images.
{\em Ribosembly} simulates compositional heterogeneity by successively adding protein subunits and ribosomal RNA, resulting in $16$ discrete structural states~\cite{qin2023cryo}. 
It has $335{,}240$ particle images, with non-uniform distribution over the 16 compositional states.
All particle images have $128\times 128$ pixels, 
% a box size of $L=128$ 
and are simulated with realistic point spread functions and a signal-to-noise ratio (SNR) of $0.01$.

{\bf Experimental Datasets.}
We also evaluate on two real datasets:
EMPIAR-10076 is a dataset comprising assemblies of intermediates of the {\em Escherichia coli} large ribosomal subunit (LSU)~\cite{davis2016modular}, with $131{,}899$ particle images ($320\times 320$ pixels, with pixel size $1.31$~{\AA}).
In the original study, four major assembly states were identified~\cite{davis2016modular}, with a subset of particles labeled as unassigned (non-ribosomal impurities) or 30S subunits.
We also consider EMPIAR-10180, a conformationally heterogeneous dataset of Pre-Catalytic Spliceosome~\cite{plaschka2017structure}.
%, composed of the U4/U6.U5 tri-snRNP associating with the U2 snRNP via the U2/U6 helix II and the U2 SF3b-containing domain.
A total of $327{,}490$ particle images were collected ($320\times 320$ pixels, with pixel size $1.69$~{\AA}).
Consistent with other heterogeneity methods considering this dataset~\cite{Recovar-2025,cryoDRGN2021} we perform analysis on a filtered subset of $139{,}722$ images.
% Previous analysis by cryoDRGN~\cite{cryoDRGN2021} identified a subset of particles producing poorly resolved maps which we exclude to get $139{,}722$ remaining images.

{\bf Implementation Details.}
% \Marcus{This paragraph could probably go to to supplemental.}
For part discovery, we seed $G\hspace{-2pt}=\hspace{-2pt}2{,}048$ components using the rigid reconstruction and adopt lightweight MLPs with a single hidden layer of $H\hspace{-2pt}=\hspace{-2pt}32$ units.
The latent space, $\zSpace$, has dimensionality $D\hspace{-2pt}=\hspace{-2pt}4$ and the feature space, $\fSpace$, has dimensionality $E\hspace{-2pt}=\hspace{-2pt}24$.
We optimize the part discovery model for $15$ and $50$ epochs on synthetic and experimental datasets.
The part-aware GMMs are optimized for $30$ epochs, using $G\hspace{-2pt}=\hspace{-2pt}8{,}192$ components, except for Ribosome synthetic and experimental datasets with $G\hspace{-2pt}=\hspace{-2pt}16{,}384$, and have MLPs with three hidden layers and $H\hspace{-2pt}=\hspace{-2pt}128$ hidden units.
On experimental datasets, we perform part discovery on downsampled $128\times 128$ images for efficiency, while the part-aware GMM is optimized on $256\times 256$ images.
We use batch size $B\hspace{-2pt}=\hspace{-2pt}64$ and set hyperparameters $\lambda_z=0.1, \lambda_f = 0.01$.
The optimization runs on a single NVIDIA GeForce RTX 2080, taking 3 to 6 hours depending on the number of Gaussians in the model. 

{\bf Evaluation Metrics.}
The quality of cryo-EM density maps are evaluated using Fourier Shell Correlation (FSC)~\cite{van2005fourier}, which is the normalized cross-correlation between two independently estimated density maps, as a function of frequency.
Metrics for heterogeneity are less standardized, but the most common is Per-Conformation FSC (or Per-Conf FSC)~\cite{jeon2024cryobench}, proposed by CryoBench \cite{jeon2024cryobench}.
Per-Conf FSC is the average FSC between the ground-truth 3D structure of a particle state, and the 3D structure corresponding to the average latent position of images associated with that state.
% To compute this metric, we obtain the reconstruction corresponding to a representative latent code for each ground-truth state, computed by averaging latent coordinates of associated images.
% FSC~\cite{van2005fourier}, the gold-standard metric in cryo-EM, is then computed against the ground-truth volume (see Supplement for more details).
% For synthetic data, we use the standard $0.5$ cut-off threshold to compute the nominal resolution from the FSC curve.
% Finally, on experimental datasets, we demonstrate that our hierarchical \methodName{} framework learns to decompose the structure into meaningful parts for compositional heterogeneity, or identifies coherent structural regions consistent with domains mentioned in prior study.
The Per-Conf FSC requires knowledge of ground-truth 3D structures for each image which is not available for experimental data and we instead rely on qualitative evaluation of the estimated parts and structures.
FSC results in a curve which can be summarized by computing the area under the curve (AUC) to more easily compare methods.
See the supplement for more details on metrics.

% \DavidF{More on Fourier Shell Correlation (FSC), with reference to supplementary material for more details.}

% To quantitatively assess reconstruction quality, we adopt the Per-Conformation Fourier Shell Correlation (Per-Conf FSC) metric introduced by CryoBench~\cite{jeon2024cryobench}. 
% To compute Per-Conf FSC, we compute a representative latent code for each ground-truth state by averaging the latents codes of its associated images, followed by obtaining the corresponding reconstruction. 
% We then compute FSC, the gold-standard metric in cryo-EM, against the ground-truth structure as the measure of reconstruction accuracy (see Supplement for more details).
% We also use the standard $0.5$ threshold to compute the nominal resolution for synthetic data.

\begin{table*}[t]
  \centering
  \footnotesize
  \resizebox{0.9\textwidth}{!}{
      \begin{tabular}{c|c|c|c|c|c|c}
        \toprule
        \multirow{2}{*}{\textbf{Method}} & \multicolumn{2}{|c|}{\textbf{IgG-1D}} & \multicolumn{2}{|c|}{\textbf{IgG-RL}} & \multicolumn{2}{|c}{\textbf{Ribosembly}} \\
        \cmidrule(r){2-7}
        & Mean (std) & Med & Mean (std) & Med & Mean (std) & Med \\
        \midrule
        3D Classification~\cite{Scheres2007} & 0.297 (0.019) & 0.291 & 0.309 (0.01) & 0.307 & 0.289 (0.081) & 0.288 \\
        CryoDRGN~\cite{cryoDRGN2021} & 0.366 (0.003) & 0.366 & 0.349 (0.008) & 0.348 & 0.415 (0.019) & 0.415 \\
        CryoDRGN-AI-fixed~\cite{levy2024revealing} & 0.366 (0.001) & 0.366 & 0.355 (0.007) & 0.354 & 0.372 (0.032) & 0.374 \\
        % Opus-DSD & 0.34 (0.006) & 0.341 & 0.346 (0.029) & 0.349 & 0.372 (0.046) & 0.382 & 0.256 (0.038) & 0.266 & 0.228 (0.030) & 0.242\\
        3DFlex~\cite{3DFlex-NM-2023} & 0.336 (0.002) & 0.336 & 0.339 (0.007) & 0.339 & - & - \\
        3DVA~\cite{3DVA2021} & 0.351 (0.003) & 0.351 & 0.341 (0.006) & 0.341 & 0.375 (0.038) & 0.372 \\
        RECOVAR~\cite{Recovar-2025} & \underline{0.391 (0.001)} & \underline{0.391} & \underline{0.372 (0.008)} & \underline{0.371}  & \textbf{0.430 (0.016)} & \textbf{0.432}
        \\
        \methodName{} (ours) & \textbf{0.402 (0.002)} & \textbf{0.402} & \textbf{0.386 (0.014)} & \textbf{0.389}  & \underline{0.427 (0.014)} & \underline{0.424} \\
        \bottomrule
      \end{tabular}
    }
    \vspace{-2pt}
    \caption{
    Mean (standard deviation) and median of AUC of Per-Conformation FSCs on CryoBench datasets~\cite{jeon2024cryobench}.
    Statistics computed across different structural states, i.e. $100$ for IgG-1D and IgG-RL and $16$ for Ribosembly 
    % FSCs are computed after masking out background noise.
    (Best method in bold, second best underlined).
    }
     \vspace*{-0.2cm}
    \label{tab:tab1_fsc}
\end{table*}

\begin{figure}[t]
\centering
\includegraphics[width=0.95\linewidth]{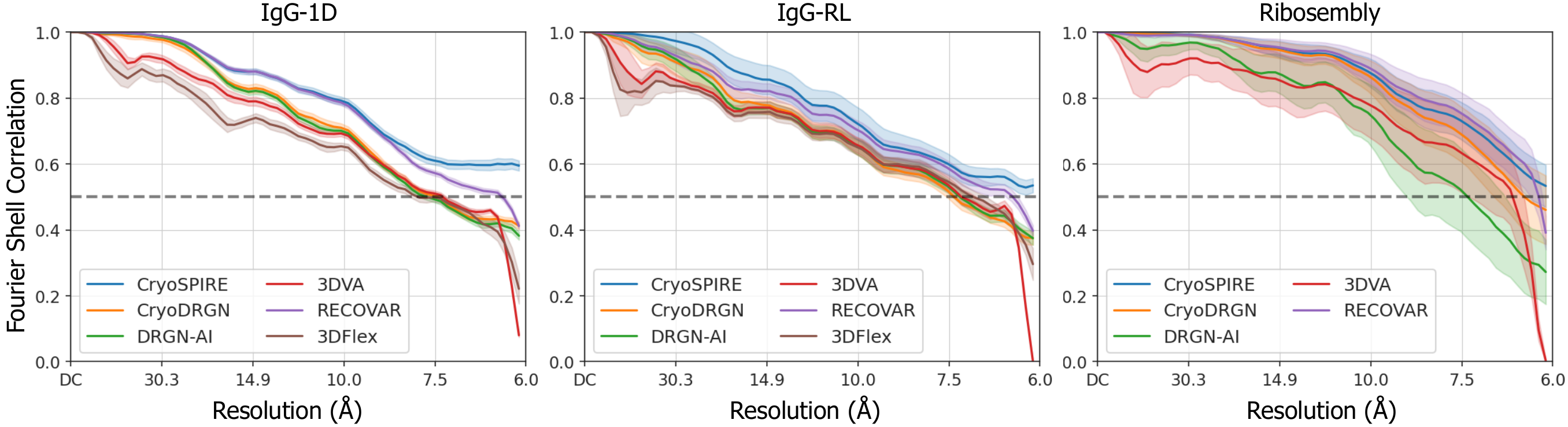}
    \vspace{-2pt}
    \caption{
    Per-Conformation FSC on CryoBench datasets.
    % of IgG-1D, IgG-RL and Ribosembly.
    Error bars indicate standard deviation across different conformations.
    The highest possible resolution is 6 {\AA} on these synthetic datasets.
    }
    \vspace*{-0.2cm}
    \label{fig:fscs}
\end{figure}

\begin{figure}[t]
    \centering
    \includegraphics[width=0.95\linewidth]{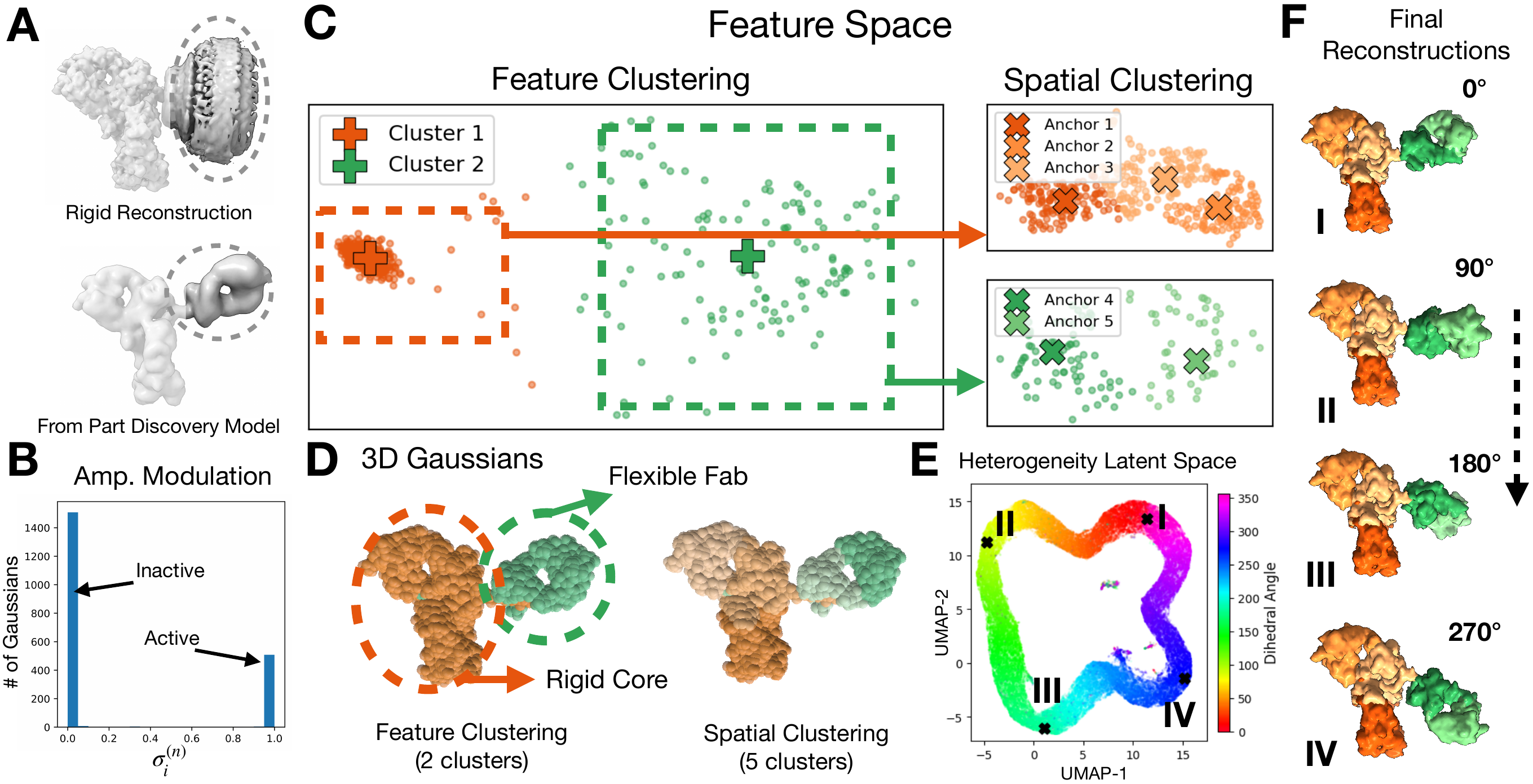}
    \vspace{-6pt}
    \caption{Results on IgG-1D~\cite{jeon2024cryobench}.
    % \Marcus{We need to label the Fab domain in the figure.  I also wonder if we should move the coarse GMM visualizations to supplemental.  I think they may be tricky to explain because to an untrained eye they may look worse than the homogenous reconstruction.}
    \textbf{(A)} Due to large motion, the Fab domain (circled) is smeared out in  rigid reconstruction, while our part discovery model identifies this domain and resolves its structure and motion, providing good initialization for subsequent modeling. 
    \textbf{(B)} For a sample structure, the histogram of amplitude modulations
    % , $m^{(n)}_i$, 
    indicate active and inactive Gaussians.
    % , with active ones yielding sharper structural features as depicted in our consensus map.
    \textbf{(C)} Gaussian feature space, $\fSpace$, shows two distinct groups (\textcolor{fab-green}{green}, \textcolor{core-orange}{orange}), corresponding to the flexible Fab  and the rigid core; feature clustering finds these groups and  divides further based on spatial proximity, yielding $5$ parts.
    \textbf{(D)}
    Configuration of 3D Gaussians after Level-1 and Level-2 clustering.
    \textbf{(E)}
    The latent space, $\zSpace$, captures conformation change (colored based on  ground truth Fab orientation).
    % , with color indicating Fab orientation , color coded with the dihedral angle, shows a circular manifold.
    \textbf{(F)}
    Sample structures from  model corresponding to four latent points, showing rotation of the Fab domain (\textcolor{fab-green}{green}).
    % where the Fab domain, comprising two parts, undergoes a full rotation relative to the rest of the complex.
    % We sample the latent space at four different dihedral angles and show the corresponding states. Samples
    % The Fab domain, covered by two segments, undergoes a $360$ motion relative to the rest of the complex.
    % \Marcus{Changes here: 
    % 1) "Coarse GMM" to "Part Discovery Model" in A.
    % 2) $\mu$ to $\sigma$ in B.
    % 3) Call it "feature clustering" and "spatial clustering" in C/D.
    % 4) Change direction of arrows in B.
    % }
    }
    \vspace*{-0.3cm}
    \label{fig:IgG-1D}
\end{figure}

\begin{figure}[t]
    \centering
    \includegraphics[width=0.975\linewidth]{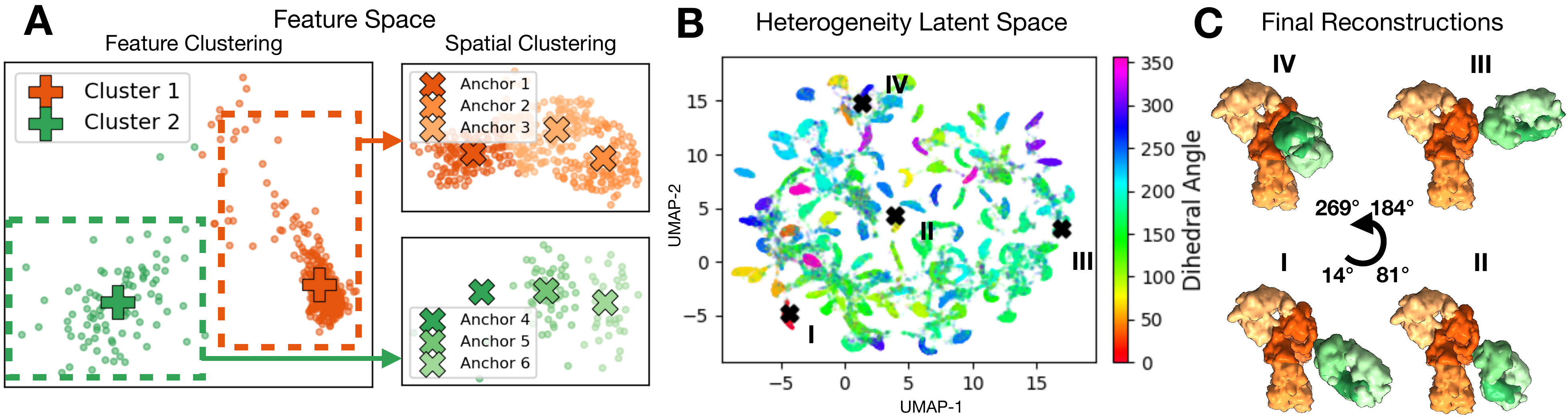}
    \vspace{-4pt}
    \caption{
    Results on IgG-RL~\cite{jeon2024cryobench}.
    \textbf{(A)} The feature space, $\fSpace$, shows two parts (\textcolor{fab-green}{green} and \textcolor{core-orange}{orange}) corresponding to the flexible Fab domain and the rigid core.
    Subsequent spatial clustering yields six parts.
    % Clustering identifies two groups which are further clustered based on spatial proximity, yielding a total of $6$ sub-segments.
    \textbf{(B)}
    The latent space, $\zSpace$, is colored with Fab orientation along with four sampled latent points that capture rotation of the Fab domain (comprising three parts).
    The motion of the Fab domain in IgG-RL is not as regular as that in IgG-1D, as reflected in the latent space.
    \textbf{(C)}
    The corresponding density maps are provided with parts illustrated in different colors.
    % \Marcus{Make the figure here consistent with the changes suggested for the IgG-1D figure.}
    }
    \vspace*{-0.5cm}
    \label{fig:IgG-RL}
\end{figure}

\begin{figure}[t]
    \centering
    \includegraphics[width=0.925\linewidth]{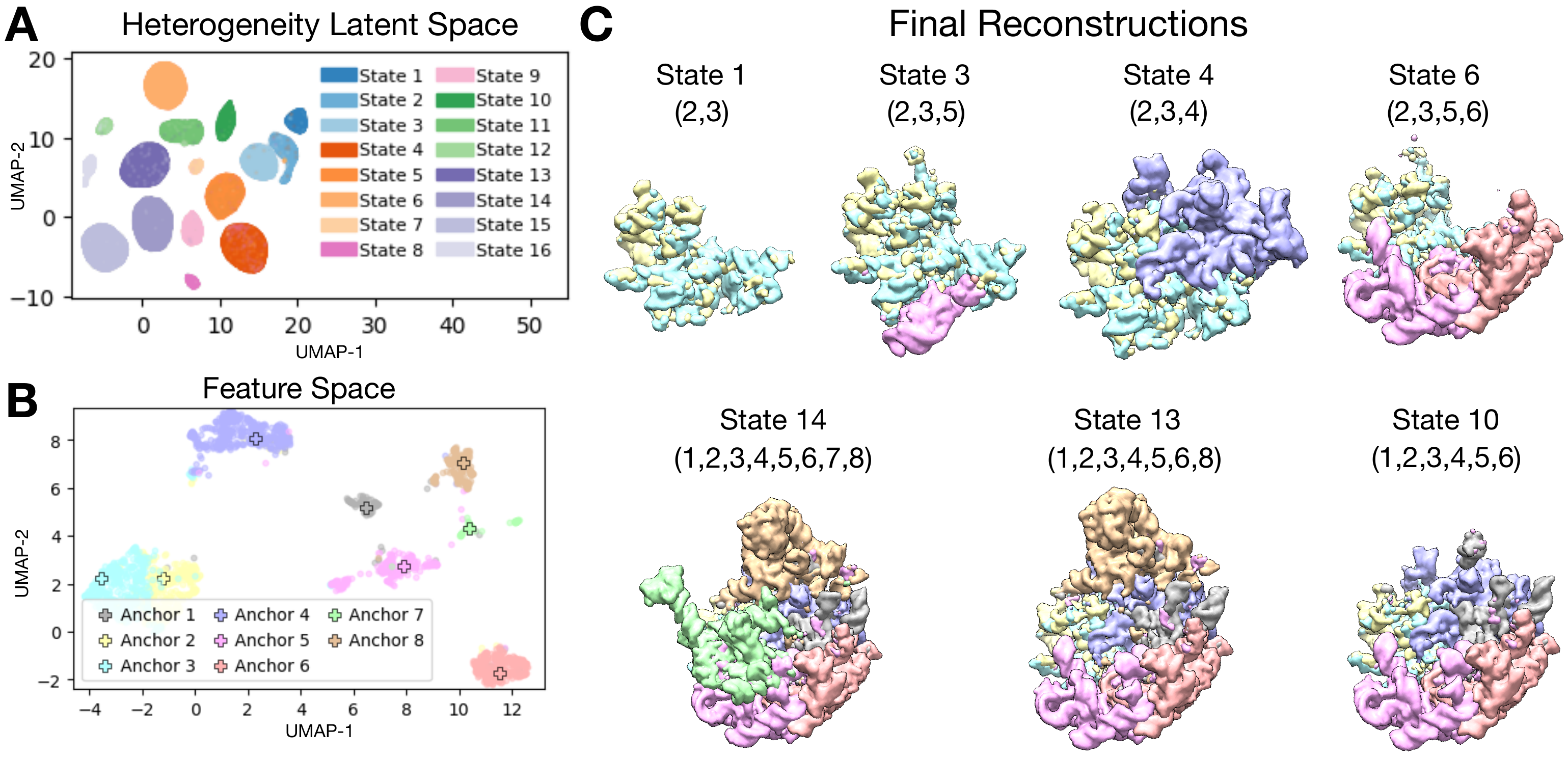}
    \vspace{-2pt}
    \caption{Results on Ribosembly ~\cite{jeon2024cryobench}
    \textbf{(A)} Gaussian feature space, $\fSpace$, showing eight major parts identified through clustering.
    \textbf{(B)} Heterogeneity latent space, $\zSpace$, colored coded with the ground-truth compositional state.
    \textbf{(C)} Visualizations of 3D density maps corresponding to seven points in latent space, with colors depicting parts (given in parentheses).
    }
    \label{fig:Ribosembly}
     \vspace*{-0.1cm}
\end{figure}

\vspace*{-0.25cm}
\section{Results}
\vspace*{-0.275cm}

% Quantitative results on three cryoBench~\cite{jeon2024cryobench} datasets, based on FSC against ground truth structures, appear in Table~\ref{tab:tab1_fsc} and 
% and Fig.~\ref{fig:fscs}.
% \begin{itemize}
%     \item 3DVA and 3DFlex are among the most widely used methods in sutrctural bio, and we grealy outperform both methods.  3DFlex is cannot cope with compositional heterogeneity and hence it is very poor on Ribosembly and only works well on smaller motions.  
%     \item LVMs, 3DVA and Cryo-DRGN, are better than 3DFlex, esp on Ribosembly, but still underperform \methodName{} by a large margin.  \item The best competing method is RECOVAR, which is a linear subspace model but it is slow and memory intensive, by comparison, and does not explicitly represent motion and, compared e.g. to 3DFlex does not give SOTA resolutions.  One can see this, in Fig 3A,B where \methodName\ FSC curves intersection the 0.5 curve at higher resolution.  While \methodName\ is significantly better than RECOVER on IgG datasets, their performance on Ribosembly are not statistically different.
%     \item First time that a GMM has been successfully applied to this benchmark (despite attempts by the authors of the benchmark).
% \end{itemize}
% \Shayan{I wrote the following based on the above suggestions:}
Quantitative comparison on the three relevant CryoBench~\cite{jeon2024cryobench} datasets are provided in Table~\ref{tab:tab1_fsc} and Fig.~\ref{fig:fscs}.
Note that \methodName{} outperforms 3DVA and 3DFlex which are among the most widely used methods in cryo-EM at present. 
As 3DFlex cannot handle compositional changes, it was not evaluated on Ribosembly.
\methodName\  outperforms non-linear latent variable models, Cryo-DRGN and DRGN-AI, especially on IgG-1D and IgG-RL by a large margin.
The most competitive method is the linear subspace model of RECOVAR, which, as reported, is memory intensive due to allocation of several bases and is not as interpretable without motion modeling. 
While \methodName{} significantly outperforms RECOVAR on IgG datasets, its performance on Ribosembly, where linear subspace models are more favorable by design, is not statistically different from RECOVAR.
Relative to the nominal FSC threshold of 0.5 for comparison to ground truth \cite{Rosenthal-Henderson-JMB2003}, the FSC curves in Fig.~\ref{fig:fscs} indicate that \methodName{} finds higher resolution density maps. 
Finally, we note that \methodName\ is the first GMM-based method to be successfully evaluated on CryoBench. 

% Quantitative results are only available for IgG-1D, IgG-RL and Ribosembly where ground truth states are available.
% The average Per-Conf FSC curves are shown in Fig.~\ref{fig:fscs} and the AUCs are reported in Tab.~\ref{tab:tab1_fsc}, demonstrating that \methodName{} achieves state-of-the-art performance.
% \DavidF{This is the main result.  Text seems a little anticlimactic. Discuss performance vs key other methods, mention downside of recovar etc.}
% We next provide a further detailed qualitative analysis for individual structures.

{\bf IgG-1D \& IgG-RL (CryoBench \cite{jeon2024cryobench}).}
The flexible Fab domain (circled in Fig.~\ref{fig:IgG-1D}A, top) in the rigid reconstruction, used as input for part discovery, is poorly resolved.
However, the part discovery model learns to selectively deactivate incoherent parts, as shown in the histogram of the modulation factors $\sigma_i^{(n)}$ in Fig.\ref{fig:IgG-1D}B.
This enables a more robust initialization (Fig.~\ref{fig:IgG-1D}A, bottom) of the hierarchical GMM.
% This enables an improved consensus map (Fig.~\ref{fig:IgG-1D}, A, bottom) to robustly initialize our hierarchical model.
The Gaussian feature space, $\fSpace$, shows two clusters corresponding to the flexible Fab domain from the rigid core (Fig.~\ref{fig:IgG-1D}C for IgG-1D and Fig.~\ref{fig:IgG-RL}B for IgG-RL).
Spatial clustering produces a total of five and six anchors for IgG-1D and IgG-RL, respectively.
The latent heterogeneity space, $\zSpace$, indicates a circular manifold of dihedral angles for IgG-1D, see Fig.~\ref{fig:IgG-1D}D. 
% \Marcus{It's not obvious to me from looking at this figure.}
% For the more challenging IgG-RL, the images form groups with dihedral similarity (Fig.~\ref{fig:IgG-RL}, B).
Four structures from the latent space in both datasets demonstrate that the Fab domain, covered by a few parts, undergo a large, predominantly rigid motion, while the rest of the complex is remains fixed.
% \DavidF{explain why conformational landscape in IgD-1D is relatively simple, but IgG-RL is much more complex?}
% The average Per-Conf FSC curves (see Fig.~\ref{fig:fscs}) and AUC (see Tab.~\ref{tab:tab1_fsc}), shows that \methodName{} achieves state of the art results.
% Also, the statistics of area under the FSC curves (AUC-FSC) computed across different conformational states is reported in Table~\ref{tab:tab1_fsc}, demonstrating that \methodName\ achieves the state-of-the-art performance.

{\bf Ribosembly (CryoBench~\cite{jeon2024cryobench}).}
After part discovery, we obtain eight parts (see Fig.~\ref{fig:Ribosembly}A) that are used to initialize eight anchors in the part-aware GMM.
% The hierarchical models is thus initialized with eight anchors.
In Fig.~\ref{fig:Ribosembly}B the learned latent space, $\zSpace$, clearly distinguishes between the different compositional states.
% For each state, we compute corresponding density maps segmented based on the learned Gaussian embedding.
For seven selected states, we visualize the estimated structure (Fig.~\ref{fig:Ribosembly}C), colored based on the discovered parts.
% The visualization in Fig.~\ref{fig:Ribosembly}, C, shows how each state is constructed by composition of common parts.
% Finally, we visualize the average Per-Conf FSC curve over all $16$ states with error bars in Fig.~\ref{fig:fscs}, right.
% Corresponding quantitative results are provided in Table~\ref{tab:tab1_fsc}, demonstrating that our method performs on par with RECOVAR while surpassing other approaches.

% \subsection{Experimental (Real) Datasets}
\begin{figure}[t]
    \centering
    \includegraphics[width=0.95\linewidth]{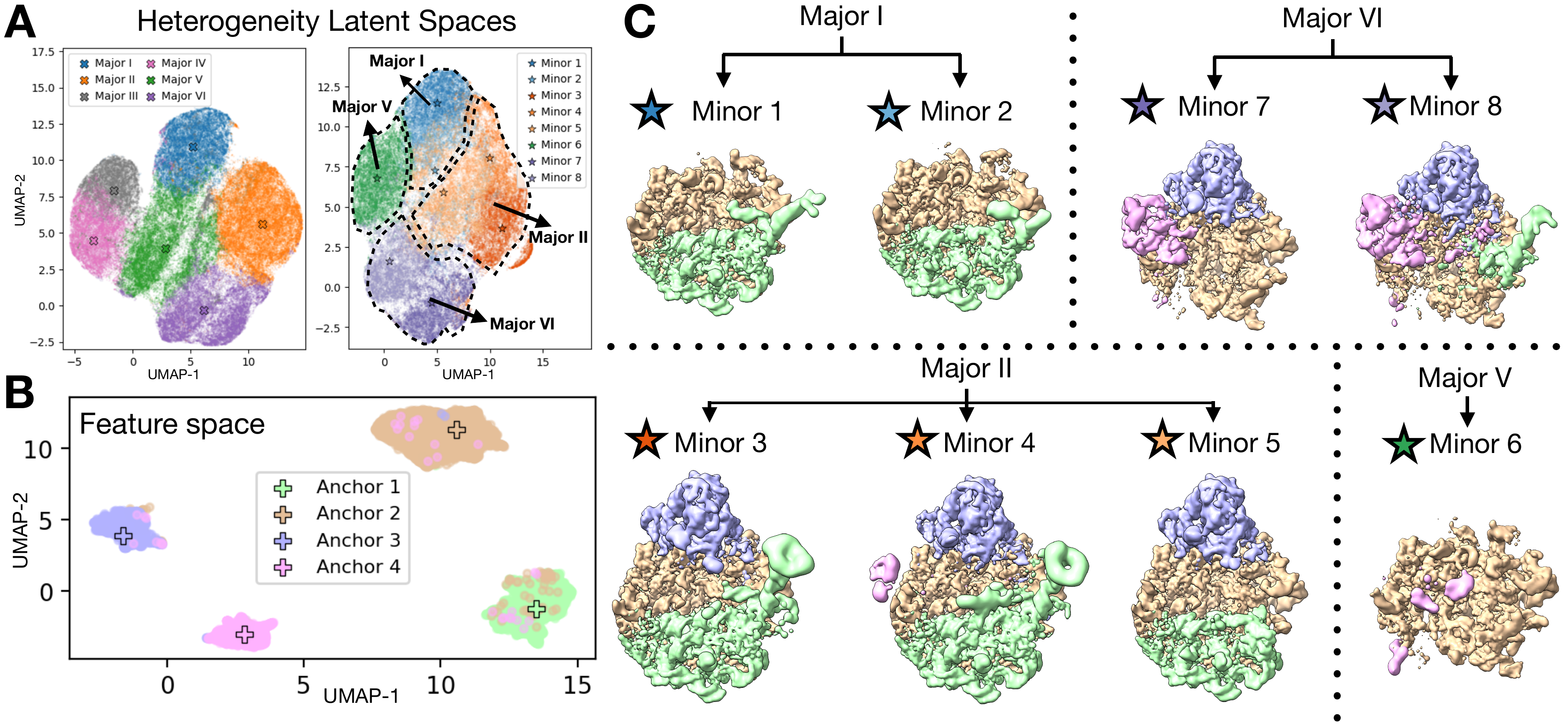}
    \vspace{-2pt}
    \caption{
    Results on Large Ribosomal Subunit (EMPIAR-10076~\cite{davis2016modular}).
    \textbf{(A)} The learned heterogenity latent spaces, $\zSpace$, in part discovery (left) identifies the four major assembly states (I, II, V, VI) and two groups of impurities (III, IV).
    After fitting the part-aware model, the major states, with impurities excluded, can be further categorized (right) into eight color-coded minor structural states.
    % Some minor states group together and represent a major state identified by the coarse-grained model, as indicated by the area enclosed by dashed lines.
    \textbf{(B)} The part discovery Gaussian feature space, $\fSpace$, reveals four parts which are used to construct the part-aware model.  
    % Using clustering, we infer the segmentation and initialize hierarchical model with four parts.
    \textbf{(C)} The structures corresponding to different states, colored by inferred part.
    % After sampling the center latent coordinates of minor clusters, we visualize corresponding segmented density maps showing how each state is generated by composing four inferred parts.
    % \textbf{(A)} UMAP visualization of the coarse-GMM latent space, color-coded with six Gaussian mixture components, producing four major assembly states labeled as I, II, V, and VI.
    % Contaminants are assigned to states III and IV. 
    % \textbf{(B)} UMAP visualization of Gaussian embeddings. Four segments are obtained using K-means++.
    % \textbf{(C)} Reconstruction of each major state is composition of four identified segments color-coded accordingly; e.g. segment 3 is the core that is included in all states while segment 1 is only in major states I and II.
    % \textbf{(D)} UMAP visualization of the latent space of Scaffold-GMM, optimized on filtered dataset. 
    % Through clustering, eight minor states are identified.
    % The area corresponding to each major state is enclosed by dashed lines.
    % \textbf{(E)}
    % Reconstructions of corresponding minor states are provided, color-coded based on underlying learned hierarchy.
    % \Marcus{Gaussian feature space is somewhat important.  It supports effectiveness of part discovery approach.  We may need to do it for space, but lets try to shrink some other stuff first.  E.G., the quantitative results on cryobench could be one figure.  But the density of the coarse GMM should probably not be shown.}
    }
     \vspace*{-0.2cm}
    \label{fig:10076-result}
\end{figure}

{\bf Large Ribosomal Subunit (EMPIAR-10076~\cite{davis2016modular}).}
% After part discovery, we find eight components in the latent space (Fig.~\ref{fig:10076-result}A) from which we can identify four major assembly states denoted as (I, II, V, VI) matching classes (C, E, B, D) from the original study~\cite{davis2016modular}, while unassigned particles and 30S contaminants are grouped in states III and IV.
We find four major assembly states in the part discovery latent space (labeled as (I, II, V, VI) in Fig.~\ref{fig:10076-result}A, left), which match classes (C, E, B, D) in the original study~\cite{davis2016modular}, with unassigned particles and 30S contaminants grouped in states III and IV, which are excluded when optimizing hierarchical model (See supplement for more details).
The Gaussian feature space, $\fSpace$, (Fig.~\ref{fig:10076-result}B) shows four distinct parts which also align with previously reported structural blocks in the original study (cf.~\cite{davis2016modular}, Fig.~6).
By analyzing the heterogeneity latent space, $\zSpace$, of the part-aware model (Fig.~\ref{fig:10076-result}A, right), we show that the major states can be further divided into subpopulations; e.g., the major state I is represented with minor states (1, 2) and the major state II has branched into minor states (3, 4, 5).
The associated structures, shown in Fig.~\ref{fig:10076-result}E, are consistent with minor states reported in the original study~\cite{davis2016modular}.
% After part discovery, we find eight minor assembly states in the latent space (Fig.~\ref{fig:10076-result}A), which form four major clusters, denoted as (I, II, V, VI), based on the latent space of coarse GMM, matching classes (C, E, B, D) from the original study~\cite{davis2016modular} (see the supplement).
% while unassigned particles and 30S contaminants are grouped in states III and IV.
% Visualizing the Gaussian feature embeddings (Fig.~\ref{fig:10076-result}B) reveals four distinct parts found by k-means++.

% Fig.~\ref{fig:10076-result}C shows the segmented coarse density maps of the major states, showing composition of parts to construct each state.
% ; e.g. the major state I comprises parts (1, 2), whereas state VI includes parts (2, 3, 4).
% Notably, these learned parts closely align with previously reported structural blocks in the original study (cf.~\cite{davis2016modular}, Fig.~6).
% We initialize \methodName{} with four anchors and find eight components in the resulting latent coordinate space, corresponding to eight distinct minor states.
% Fig.~\ref{fig:10076-result}D, shows that the previously obtained major states could be further divided into refined minor subpopulations; e.g. the major state I is represented with minor states (1, 2) and the major state II has branched into minor states (3, 4, 5).
% Associated higher-resolution density maps are demonstrated in Fig.~\ref{fig:10076-result}E, consistent with minor states reported in the original study~\cite{davis2016modular}.

\begin{figure}[t]
    \centering
    \includegraphics[width=0.925\linewidth]{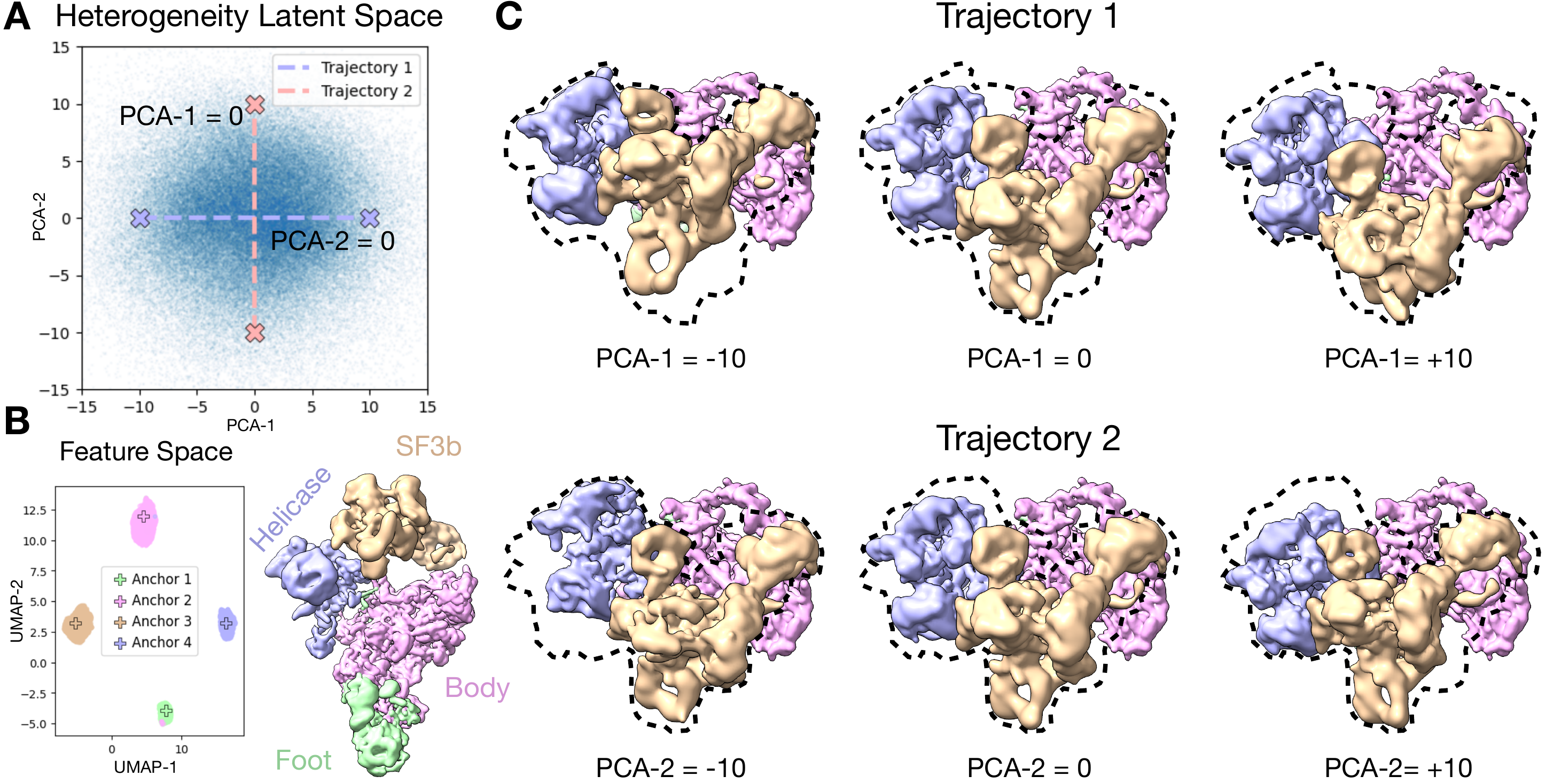}
    \caption{Results on Pre-Catalytic Spliceosome (EMPIAR-10180~\cite{plaschka2017structure}).
    \textbf{(A)} PCA of the latent space, $\zSpace$, is used to generate two structural trajectories.
    \textbf{(B)} The Gaussian feature space, $\fSpace$, shows four parts which correspond to known helicase, SF3b, body and foot domains as shown in 3D visualization of Gaussian components configuration.
    \textbf{(C)} Three states along each trajectory.
    In both trajectories, body is rigid while SF3b and helicase show large-scale motion.
    % \Marcus{Can we generate the plot A with square axes?  Right now it makes it look a little like the latent space is isotropic which would make the two principal directions somewhat arbitrary.  Not critical.}
    % , along with subtle bending or motion of the foot.
    % \Shayan{Motion of the foot is not visible from top view.}
    % \textbf{(D)} Front and side views of the reconstruction corresponding to origin of latent space based on the Scaffold-GMM. Higher resolution details are resolved.
    % \textbf{(E)} We identify similar trajectories in the latent space using PCA and focus on the motion of SF3b and helicase domains using the top view.
    % Four reconstructions along each trajectory is selected and visualized. Corresponding latent coordinates is written below each reconstruction. 
    % For Trajectory 1, $\text{PCA-2}=0$. For Trajectory 2, $\text{PCA-1}=0$.
    % \Shayan{Caption needs to be revised as the figure has changed.}
    }
     \vspace*{-0.2cm}
    \label{fig:10180-result}
\end{figure}

{\bf Pre-Catalytic Spliceosome (EMPIAR-10180 \cite{plaschka2017structure}).}
The feature space, $\fSpace$, of the part discovery model (Fig.~\ref{fig:10180-result}B), shows four distinct clusters, which correspond to coherent structural regions -- foot, body, helicase, and SF3b -- consistent with the original study~\cite{plaschka2017structure}.
Accordingly, we optimize the part-aware model with four anchors.
To illustrate structural variability, we run PCA on the heterogeneity latent space, $\zSpace$, and extract two principal directions illustrated in Fig.~\ref{fig:10180-result}A.
Top views of density maps along the two principal directions (Fig.~\ref{fig:10180-result}C) show two modes of conformational heterogeneity.
The first direction reflects a forward–backward rotation of the SF3b and helicase regions.
The second direction captures a side-to-side rotation of SF3b, and a diagonal shift of the helicase.
Please see the supplement for more visualization on conformational changes.
% , along with a slight bending of the foot domain.
% , and a subtle lateral motion of the foot.
% \Marcus{Higher resolution than what?}
% With \methodName{} we obtain higher-resolution reconstruction, as shown in Fig.~\ref{fig:10180-result}D.
% We visualize top views of corresponding reconstruction sampled from the trajectories, as in Fig.~\ref{fig:10180-result}E, focusing on the local motion SF3b and helicase domains.

\begin{figure}[t]
    \centering
    % Creating a row with two columns using minipage
    \begin{minipage}{0.52\textwidth} % Adjust width as needed
    \centering
      % \footnotesize
      \resizebox{1\textwidth}{!}{
          \begin{tabular}{c|c|c|c}
            \toprule
            \textbf{Method} & \textbf{IgG-1D} & \textbf{IgG-RL} & \textbf{Ribosembly} \\
            \midrule
            \methodName{} & \textbf{0.402 (0.002)} & \textbf{0.386 (0.014)} & \textbf{0.427 (0.014)} \\
            w/o hier.\ motion & 0.388 (0.002) & 0.372 (0.010) & 0.425 (0.015) 
            \\
            over-segment  & 0.384 (0.002) & 0.375 (0.010) & 0.423 (0.016) \\
            w/ pos.\ encoding & 0.377 (0.002) & 0.361 (0.007) & 0.415 (0.023) \\
            \bottomrule
          \end{tabular}
        }
        \vspace{-5pt}
        \captionof{table}{Mean AUC-FSCs reported on datasets from CryoBench~\cite{jeon2024cryobench} for ablation study.}
    \end{minipage}%
    \hspace{0.02\textwidth}
    \begin{minipage}{0.45\textwidth}
        \centering
        \includegraphics[width=0.9\linewidth]{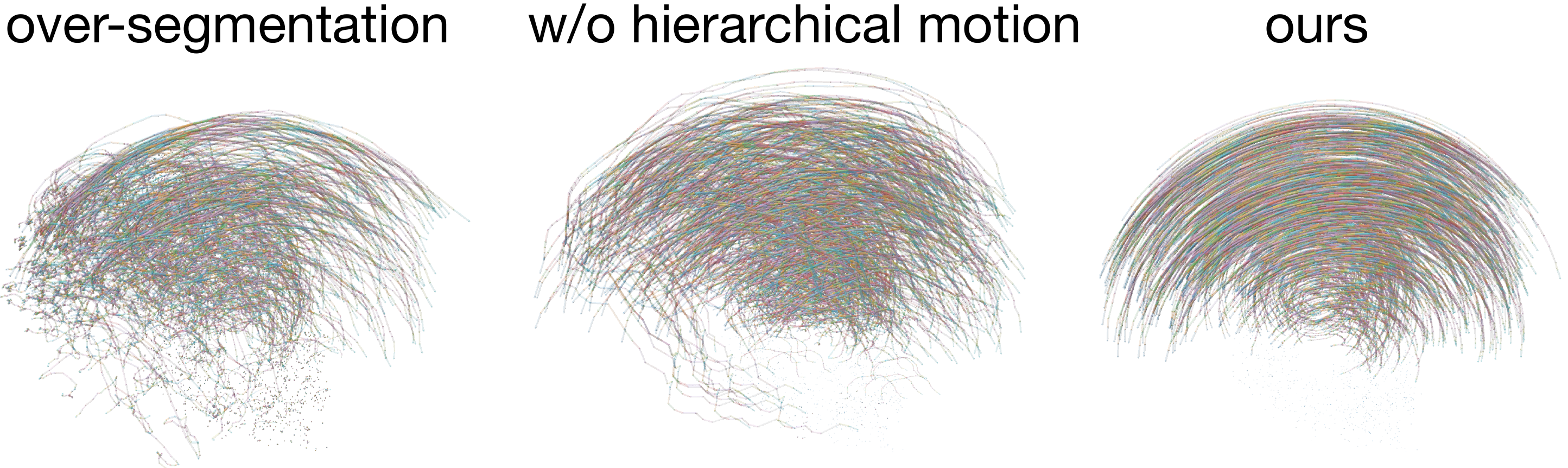}
        \vspace*{-0.2cm}
        \captionof{figure}{Estimated motion of Gaussians for $30$ states of IgG-1D. 
        The baselines fail to capture local rigidity. }
        \label{fig:gmm-trajectory}
    \end{minipage}
    \vspace*{-0.25cm}
\end{figure}

\vspace*{-0.2cm}
\subsection{Ablation Study}
\vspace*{-0.25cm}

Here, we ablate key design decisions in our framework.
% are critical to achieve the state-of-the-art results.
To demonstrate the importance of anchor-based motion modeling in \methodName{}, we consider a baseline without anchors that uses an MLP to directly learn deformations of individual Gaussians.
Quantitative comparison on IgG-1D and IgG-RL, as in Table~\ref{tab:tab1_fsc}, shows that the lack of anchor based motion leads to inferior results.
This is less critical for Ribosembly with minor conformational changes.
We also compare with a model that over-segments the structure by using $K=64$ anchors, which achieves worse performance.
In Fig.~\ref{fig:gmm-trajectory}, we visualize the estimated motion of Gaussians on the IgG-1D dataset.
Both baselines fail to capture the locally rigid and smooth motion.
Finally, we consider a baseline where the Gaussian feature space is replaced with a positional encoding, similar to previous methods, e.g., \cite{DynaMight-2024}.
This baseline is unable to identify meaningful parts and achieves inferior quantitative performance.
% , see Table~\ref{tab:tab1_fsc}.
% Finally, previous methods, e.g., \cite{DynaMight-2024}, used positional encoding~\cite{MildenhallEtAl2022} of Gaussian centers to condition the neural network.
% We consider a similar baseline where the Gaussian feature space is replaced with a positional encoding; but the method is unable to identify meaningful parts and quantitative results are degraded, see Table~\ref{tab:tab1_fsc}.

\vspace*{-0.1cm}
\section{Conclusion}
\vspace*{-0.2cm}

We present \methodName{}, a hierarchical cryo-EM density model to capture conformational and compositional heterogeneity in the 3D structure of biomolecules from 2D images.
This includes a novel method for part discovery and a hierarchical Gaussian mixture model for which the parts provide meaningful inductive biases to regularize model fitting.
\methodName{} establishes a new state-of-the-art on the CryoBench heterogeneous benchmark, and produces meaningful parts
%with high-quality 3D density maps
on experimental data.
% structures -
% We achieve state-of-the-art performance and demonstrate that \methodName{} is able to identify coherent parts from experimental dataset.
% , which was only possible through rigorous study before.
% \Shayan{future direction... interpretation of latent space in terms of free energy?}
% \Shayan{Limitations...}

While \methodName{} shows promising results, limitations exist.
Validation of estimated structures and variability from heterogeneous experimental data remains an open problem for all methods, including \methodName{}.
Interpreting the inferred latent space remains challenging, specifically how it may relate to the biophysical energy landscape of molecular states.
Finally, like other methods, we presume an initial estimate of the structure and image poses; inaccuracies in these may limit \methodName{}'s efficacy.
A fully {\em ab initio} method for heterogeneous data remains future work.

% A more compact version:
% While \methodName{} shows promising results, validating estimated structures for all methods remains an open challenge.
% Interpreting the inferred latent space in relation to the biophysical energy landscape is also underexplored.
% We also presume an initial estimate of the structue and image poses, and we will explore a fully {\em ab initio} version in future.

\section*{Broader Impact}
\vspace*{-0.25cm}
Cryo-electron microscopy (cryo-EM) has emerged as a revolutionary technique in structural biology, enabling the determination of macromolecular structures with significant societal impact. 
Computational methods, grounded in machine learning and computer vision have now been used to determine many thousands of biological structures.
Notably, cryo-EM played a pivotal role in elucidating the structure of the SARS-CoV-2 spike protein, revealing its pre-fusion conformation and aiding in the assessment of medical countermeasures. 
Complementing computational methods such as AlphaFold for protein structure prediction, cryo-EM has revolutionized our understanding of cellular processes and accelerated the development of novel therapeutics, including synthetic antibodies.
Nevertheless, we strongly condemn any usage of our proposed hierarchical 3D GMM representation for generating malicious data, improperly modifying signals, or spreading misinformation.
\begin{ack}
\vspace*{-0.2cm}
This research was supported in part by the Province of Ontario, the Government of Canada, through NSERC, CIFAR, and the Canada First Research Excellence Fund for the Vision: Science to Applications (VISTA) programme, and by companies sponsoring the Vector Institute.
\end{ack}

% \clearpage
% \bibliographystyle{unsrt}
\bibliographystyle{plain}
\bibliography{refs}

%%%%%%%%%%%%%%%%%%%%%%%%%%%%%%%%%%%%%%%%%%%%%%%%%%%%%%%%%%%%
\newpage
\appendix

\begin{center}
{\Large Supplementary Material} \\  [0.5em]
{\Large
{\bf Reconstructing Heterogeneous Biomolecules via \\ [0.2em]
Hierarchical Gaussian Mixtures and Part Discovery}} \\

\end{center}

\section{Project Webpage}
We share a \href{https://shekshaa.github.io/CryoSPIRE}{Project Webpage}, presenting short videos describing \methodName{} and showing qualitative comparisons with baseline methods on synthetic and experimental data.
We use ChimeraX~\cite{pettersen2021ucsf} to create detailed 3D visualizations of reconstructions shown in Figs.~\ref{fig:all-methods-synthetic},~\ref{fig:more-10076}, and~\ref{fig:all-methods-10180}. 
In the webpage, we also demonstrate how navigating the learned latent space leads to various structural states reconstructed by \methodName{}.

\section{GMM Image Formation, Parameterization and Rendering}
\label{sec:supp-gmm-render}
% \vspace*{-0.2cm}

% \DavidF{Rather than describe the image formation from scratch, we should ties this into the model described in the paper. Ie "We assume a model like that described in Sec XXX.  That said, we use a slightly unconventional parameterization, namely ....   \\
% More specifically, I would start by saying, we use the model in Sec XXX in the paper, but with a slightly different parameterization of the 2D projection, which we find is easier to train. ....  In other words, I'd start with the different parameterization of 2D projection, and then you just need to explain why the parameterization is really more or less just a slightly different expression for a 3D Gaussian mixture.We use this one because we find it trains more easily, even though it is mathematically the equivalent \\
% You want to avoid what looks like a new derivation of a different formulation that should have been in the paper in the first place.}
In cryo-EM, the image formation process follows integral projection of a 3D density to the 2D image plane.  
For a 3D Gaussian mixture, the projection is analytically tractable, as described in Sec.~\ref{sec:related-work}, Eq.~\ref{eq:render}.
For the purposes of optimization, however, we adopt a slightly different parameterization as we find it to be somewhat better behaved.
Assuming each 3D Gaussian component in the mixture is parameterized with center $\boldsymbol{c} \in \mathbb{R}^3$, isotropic scale $s \in \mathbb{R}$, and an amplitude $m \in \mathbb{R}$,
% we first transform the GMM into the observation space using the rotation $\boldsymbol{R} \in SO(3)$ and translation $\boldsymbol{t}  \in \mathbb{R}^3$, and then 
we define the 2D noise-free projection along the canonical $z$-axis, for location $\tilde{\boldsymbol{p}} \in \mathbb{R}^2$, as
\begin{equation}
    \tilde{I}(\tilde{\boldsymbol{p}}) ~=~ \sum_{i}  m_i \exp\left(-\frac{|| \,\tilde{\boldsymbol{p}} - [\boldsymbol{c}_i]_{xy} \, ||^2_2}{2s_i^2}\right) \ 
    \label{eq:supp-render} \
\end{equation}
Here, we modify the weight of terms such that the peak intensity of each Gaussian term solely depends on the amplitudes $m_i$, whereas, in Eq.~\ref{eq:render}, it is proportional to both $s_i$ and $m_i$.  
This leads to a direction of ambiguity in the optimization landscape where increasing one parameter (e.g. $m_i$) and decreasing the other (e.g $s_i$) can compensate; the coupling between $m_i$ and $s_i$ makes it challenging to set learning rates and it destabilizes the optimization dynamics.
The alternative parameterization in Eq.~\ref{eq:supp-render} that we use is mathematically equivalent to the following weighting of the 3D mixture:
\begin{equation}
    f(\boldsymbol{p}) ~=~ \sum_{i} \dfrac{m_i}{\sqrt{2\pi}s_i} \exp\left(-\frac{||\boldsymbol{p} - \boldsymbol{c}_i||^2_2}{2s_i^2}\right) \ , 
    \label{eq:supp-density-map-ours}
\end{equation}
for location $\boldsymbol{p} \in \mathbb{R}^{3}$. 
This simply involves a change of variables, i.e., $m_i \rightarrow m_i / \sqrt{2\pi}s_i$.

As discussed in the introduction of the paper, cryo-EM images are extremely noisy, since low electron dosages are used to minimize radiation damage to the particles.
To ensure sufficient image contrast, the microscope is defocused, which is modeled as  convolution with a image-specific point-spread function (PSF), $g^{(n)}$, or, more commonly as modulation in the Fourier domain with a contrast transfer function (CTF)~\cite{singer2020computational}.
Finally, we model all sources of noise with additive, zero-mean Gaussian noise, $\epsilon^{(n)} \sim {\cal N}(\boldsymbol{0}, \sigma^2I)$.
Taken together, the final image can be expressed as,
\begin{equation}
    \hat{I}^{(n)} \, = \, g^{(n)} \star \tilde{I}^{(n)} + \epsilon^{(n)}   
\end{equation}
For optimization, we minimize a squared L2 reconstruction loss between model predictions and observed image, which is proportional to the negative log-likelihood,
\begin{equation}
    \mathcal{L}(I^{(n)}, \hat{I}^{(n)}) \, =\,  ||I^{(n)} - \hat{I}^{(n)}||^{2}_2 \ .
\end{equation}

\paragraph{Implementation.}
Among the Gaussian parameters, the scale is constrained to be positive, and amplitudes are constrained to be non-negative.
The centers have no hard constraints.
To realize such constraints, we define $s_i$ in the log-scale domain, $s_i = \exp (\tilde{s}_i)$.
We also use a ReLU activation function to ensure that amplitudes are non-negative, i.e., $m_i = \text{ReLU}(\tilde{m}_i)$.
Thus, the free parameters are $\{(\boldsymbol{c}_i, \tilde{s}_i, \tilde{m}_i)\}_i$.

The rendering equation (Eq.~\ref{eq:supp-render}) defines a function $I$ on the  2D plane.
We discretize the function within a box $[-0.5,0.5]^2$ using a regular $L \times L$ grid, denoted $\Lambda = \{(x_u, y_v)\}_{u,v=1}^{L}$.
A naive way to evaluate $I$ on is to compute the contribution of Gaussian terms separately per-point on the grid, which can be stored collectively in a $G \times L \times L$ matrix.
Following~\cite{MuyuanChen-GMM-2023b}, we can simplify that computation since 2D Gaussians can be expressed as a separable product of 1D Gaussians (on rows and columns):
\begin{equation}
    \exp\left(-\frac{||\boldsymbol{p}_{xy} - \tilde{\boldsymbol{c}}_{i,xy}||^2_2}{2s_i^2}\right) \, =\, \exp\left(-\frac{||\boldsymbol{p}_x - \tilde{\boldsymbol{c}}_{i,x}||^2_2}{2s_i^2}\right)
    \exp\left(-\frac{||\boldsymbol{p}_y - \tilde{\boldsymbol{c}}_{i,y}||^2_2}{2s_i^2}\right) \ .
\end{equation}
All points within a single row or column in the grid share corresponding 1D Gaussian terms, which thus need to be computed only once.
To this end, we decompose the 2D grid $\Lambda$ into two 1D grids $\Lambda_x = \{x_u\}_{u=1}^{L}$ and $\Lambda_y = \{y_v\}_{v=1}^{L}$, and accordingly compute two $G \times L$ matrices, $M_x$ and $M_y$,
\begin{equation}
    M_x(i, x_u) \, =\, \exp\left(-\frac{||x_u - \tilde{\boldsymbol{c}}_{i,x}||^2_2}{2s_i^2}\right) \ , \quad  
    M_y(i, y_v) \, =\, \exp\left(-\frac{||y_v - \tilde{\boldsymbol{c}}_{i,y_v}||^2_2}{2s_i^2}\right) \ .
\end{equation}
These matrices store the value of 1D Gaussian terms on 1D grids. 

We also define all amplitudes within a $G \times 1$ matrix, denoted as $W$, with $W_{i,1} = m_i$.
Next, we use fast matrix operations on $W$, $M_x$ and $M_y$, to realize the above rendering equation (Eq.~\ref{eq:supp-render}).
We first compute vectorized outer product between $W$ and $M_x$, yielding a new $G \times 1 \times L$ matrix.
We then compute vectorized outer product between the resulting matrix and $M_y$, yielding a final matrix of size $G \times 1 \times L \times L$, followed by reducing the first dimension using summation.
As a result, we obtain a $1 \times L \times L$ matrix which is our desired discretized projection.
Importantly, compared to the naive process, this approach avoids redundant computations using separability of 2D Gaussians.

\section{\methodName{} Initialization}
\label{sec:supp-init}
% \vspace*{-0.2cm}
% \DavidF{where do CTF/PSF comes from?\\
% Where does initial consensus map come from and how critical is it?  \\
% How do we initialize the coarse-grain model for part discovery.\\
% More on optimization details?}

Before describing the initialization process of the coarse-grained GMM for part discovery, we outline how to obtain a preliminary density map as a rigid reconstruction, as well as per-image poses and CTF parameters.
For CryoBench~\cite{jeon2024cryobench} datasets, the ground-truth CTFs and poses are provided for each particle image.  
Given the poses, we use a simple form of backprojection to compute an initial rigid reconstruction.
For Ribosome experimental data~\cite{davis2016modular} (EMPIAR-10076), we estimate per-particle poses and an initial rigid reconstruction using CryoSPARC~\cite{cryoSPARC-NM-2017}.
For Spliceosome~\cite{plaschka2017structure} (EMPIAR-10180), an initial rigid reconstruction and per-particle poses are available (computed using RELION~\cite{Scheres-JSB-2012}).
For both experimental datasets, CTFs are estimated in a standard preprocessing stage.

As demonstrated in the Gaussian Splatting literature \cite{Kerbl2023tdgs}, the optimization dynamics are highly sensitive to the initial values of Gaussian parameters.
For the coarse-grained GMM used for part discovery, we use the input rigid reconstruction to seed $G=2048$ Gaussians.
Given the input density map, we discard voxels with density below a user-defined threshold, and then sample $G$ of the remaining voxels with probability proportional to density.
Gaussians are then seeded at those positions.
Note that, since flexible regions often exhibit lower density in the rigid reconstruction, setting the threshold too high may exclude these regions, resulting in poor initialization.
% To address this, we developed a visualization tool that enables users to explore and select an appropriate threshold interactively.
Finally, we initialize the amplitude and scales to user-defined values, $m_i=0.15, s_i=0.02$.

To initialize the hierarchical model, we need to determine the anchors.
We run k-means++~\cite{kmeans2007} clustering on the learned Gaussian features from the part discovery model.
UMAP visualizations (Figs.~\ref{fig:IgG-1D}~\ref{fig:IgG-RL},~\ref{fig:Ribosembly},~\ref{fig:10076-result},~\ref{fig:10180-result}) indicate that the feature space contains well-separated clusters of Gaussian components.
This clear distinction between clusters provide the potential to automatically set the number of anchors, yet for now we 
choose number of anchors manually (eg with UMAP visualization).
Each anchor receives a feature vector that is set to the centroid of its features in the cluster, and its position is set to that of the closest Gaussian component in the feature space.
Feature-based clustering is followed by spatial clustering for IgG data to further divide clusters into local regions, improving coverage of the density map.
Furthermore, the part discovery model provides an improved density map, where amplitude modulation helps to identify unused components that can be discarded.
Based on this density map, we follow a similar procedure to seed a denser set of Gaussian components.
Each new Gaussian is connected to the anchor that the closest Gaussian in the part discovery model is assigned to.
Feature offsets are initialized to zero, so Gaussians initially 
inherit the features of their anchors. 
We initialize amplitudes as $m_i=0.15$.
But we use lower initial scale values $s_i=0.01$, since this initialization uses a high quality, more reliable density map.

\section{FSC-based Performance Metrics}
\label{sec:supp-fsc}
% \vspace*{-0.2cm}

Evaluation of cryo-EM methods is very challenging, in part because ground truth 3D density maps do not exist for experimental data. 
To date, the most widely used metric, namely, Fourier Shell Correlation (or FSC), is a measure of consistency, defined as the normalized cross-correlation of two 3D density maps computed as a function of frequency~\cite{Harautz-vanHeel-1986,van2005fourier}.
Given two 3D density maps $A$ and $B$ with Fourier coefficients $A_j$ and $B_j$, the FSC at wavelength $\lambda$ is defined as
\begin{eqnarray}
{\cal F}_{\lambda} ~=~
\frac{ \sum_{j \in S_{\lambda}} A_j \, B_j^*}
{\sqrt{ \sum_{j \in S_{\lambda}} |\, A_j |^2 ~ \sum_{j \in S_{\lambda} } |\, B_j  |^2 }} ~,
\label{fsc1}
\end{eqnarray}
where $S_{\lambda}$ is the set of Fourier indicies for frequencies within a spherical shell with wavelength $\lambda$ centered at the origin.
Here,
$|z|$ is the modulus of the complex scalar $z$, and $z^*$ is the complex conjugate of $z$. 

FSC can measure consistency between an estimated density map and a ground truth map. Signal-to-noise ratios in particle images, and hence in 3D reconstructions will decrease with frequency. So FSC is usually close to 1 at low frequencies, and then decrease toward zero at higher frequencies where observations are missing or dominated by noise.
As a consequence, there are two common ways to characterize the quality of 3D reconstructions. One is the area-under-the-curve (AUC) of the FSC curve.  The other is a resolution of the map, defined as the wavelength at which FSC drops below a threshold.
A threshold of 0.5 is used for FSC computed between an estimated map and a ground truth map~\cite{Rosenthal-Henderson-JMB2003}, which 
corresponds to a SNR of 1 (an estimator of spectral SNR from FSC is simply $|FSC_\lambda | \, /\, (1-|FSC_\lambda| )$ \cite{frank_al-ali_1975,PENCZEK200234,Unser-1987}).
Since we do not generally have ground truth density maps for real experimental datasets, FSC between ground truth and estimated maps is typically only used with synthetic data, e.g, CryoBench~\cite{jeon2024cryobench}.

For homogeneous reconstruction with real experimental data, where ground truth is unavailable, FSC is used as a measure of consistency between two 'independent' 3D reconstructions.
In the ``gold standard'' protocol, one randomly divides the set of particle images into halves, from which one estimates two 3D density maps that we assume are conditionally independent given the true density.
FSC then provides a measure of consistency as a function of frequency between the two half-maps.
When using the gold standard FSC protocol (comparing two half-maps estimated from independent halves of the data), a threshold of 0.143 is used to measure the map resolution, which approximately corresponds to when the spectral SNR of a map computed from the full set of images is 1~\cite{Rosenthal-Henderson-JMB2003}. 

\begin{figure}[h]
\centering
\includegraphics[width=\linewidth]{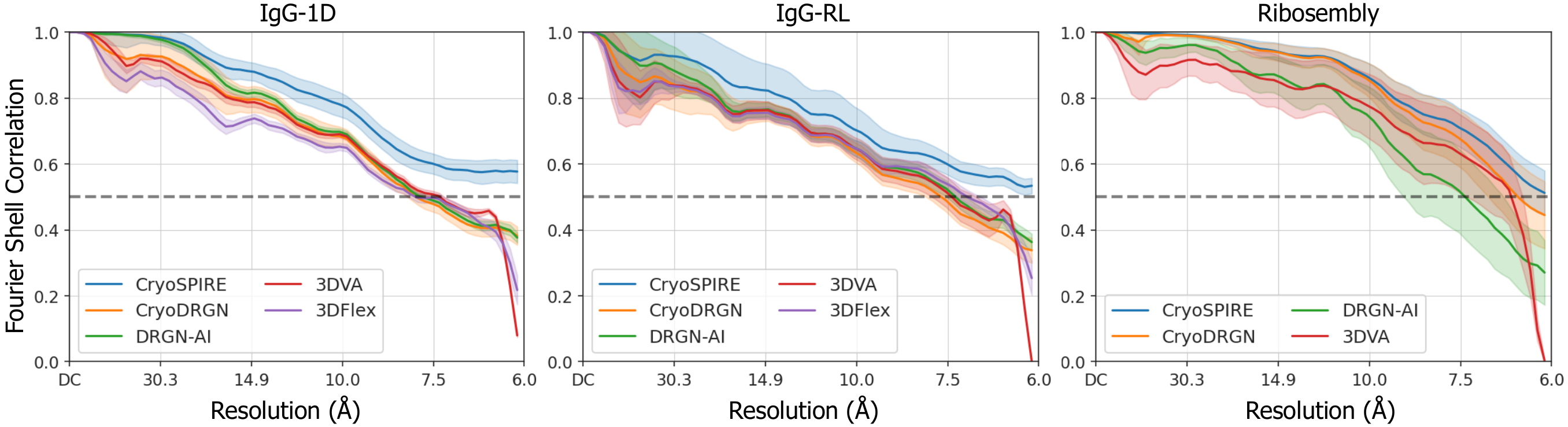}
    \vspace{-4pt}
    \caption{
    Per-Image FSCs on CryoBench datasets.
    % of IgG-1D, IgG-RL and Ribosembly.
    Error bars indicate standard deviation across different states.
    The highest possible resolution is 6 {\AA} on these synthetic datasets.
    % \Shayan{Waiting for Minkyu to share RECOVAR result.}
    }
    \vspace*{-0.3cm}
    \label{fig:per-image-fscs}
\end{figure}

Despite the widespread use of FSC with the gold standard protocol, there is no accepted extension to heterogeneous reconstruction. 
The best existing benchmark, namely CryoBench \cite{jeon2024cryobench}, leverages synthetic data, in which case one has ground truth density maps available.
To handle heterogeneity, they define the Per-Conformation FSC metric as the average AUC-FSC or FSC curve taken over all ground truth states generated in the synthetic dataset, whether compositional or conformational in its heterogeneity. 
This is the performance metric reported in Table \ref{tab:tab1_fsc} and Figure \ref{fig:fscs}.
A closely related measure is the Per-Image FSC, which is the mean AUC-FSC between the ground truth density maps and those estimated from the latent state by taking one sample particle per ground-truth state.
For completeness we report this below in Table \ref{tab:tab2_image_fsc} and visualize FSC curves in Fig.~\ref{fig:per-image-fscs}.
Due to the high memory requirements, we were unable to reproduce FSC curves for RECOVAR by the time of submission.
For more details on RECOVAR performance, see the CryoBench paper~\cite{jeon2024cryobench}.

\begin{table*}[h!]
  \centering
  \footnotesize
  \resizebox{1\textwidth}{!}{
      \begin{tabular}{c|c|c|c|c|c|c}
        \toprule
        \multirow{2}{*}{\textbf{Method}} & \multicolumn{2}{|c|}{\textbf{IgG-1D}} & \multicolumn{2}{|c|}{\textbf{IgG-RL}} & \multicolumn{2}{|c}{\textbf{Ribosembly}} \\
        \cmidrule(r){2-7}
        & Mean (std) & Med & Mean (std) & Med & Mean (std) & Med \\
        \midrule
        3D Classification~\cite{Scheres2007} & 0.297 (0.019) & 0.291 & 0.309 (0.01) & 0.307 & 0.289 (0.081) & 0.288 \\
        CryoDRGN~\cite{cryoDRGN2021} & 0.351 (0.028) & 0.356 & 0.331 (0.016) & 0.333 & 0.412 (0.023) & 0.415 \\
        CryoDRGN-AI-fixed~\cite{levy2024revealing} & 0.364 (0.002) & 0.364 & 0.348 (0.012) & 0.350 & 0.372 (0.032) & 0.375 \\
        % Opus-DSD & 0.34 (0.006) & 0.341 & 0.346 (0.029) & 0.349 & 0.372 (0.046) & 0.382 & 0.256 (0.038) & 0.266 & 0.228 (0.030) & 0.242\\
        3DFlex~\cite{3DFlex-NM-2023} & 0.335 (0.003) & 0.335 & 0.337 (0.007) & 0.337 & - & - \\
        3DVA~\cite{3DVA2021} & 0.349 (0.004) & 0.350 & 0.333 (0.014) & 0.335 & 0.375 (0.038) & 0.375 \\
        RECOVAR~\cite{Recovar-2025} & \underline{0.386 (0.001)} & \underline{0.388} & \underline{0.363 (0.011)} & \underline{0.363}  & \textbf{0.429 (0.018)} & \textbf{0.432}
        \\
        CryoSPIRE (Ours) & \textbf{0.396 (0.013)} & \textbf{0.400} & \textbf{0.375 (0.020)} & \textbf{0.391}  & \underline{0.422 (0.015)} & \underline{0.421} \\
        \bottomrule
      \end{tabular}
    }
    \caption{
   Mean AUC of {\bf Per-Image FSC} is reported for various methods on CryoBench  datasets~\cite{jeon2024cryobench}.
    In parentheses, we report the standard deviation indicating the spread of AUC among different structural states ($100$ states for IgG-1D and IgG-RL and $16$ states for Ribosembly).
    FSC curves are computed after masking out background noise.
    We AUC numbers from CryoBench~\cite{jeon2024cryobench} for RECOVAR.
    (Best method in bold, second best underlined.)
    }
    \label{tab:tab2_image_fsc}
\end{table*}

\vfill

\clearpage
\section{Complete Qualitative Result on Ribosembly}
\label{sec:supp-qual-ribosembly}
\vspace*{-0.2cm}

\begin{figure}[h!]
    \centering
    \includegraphics[width=0.85\linewidth]{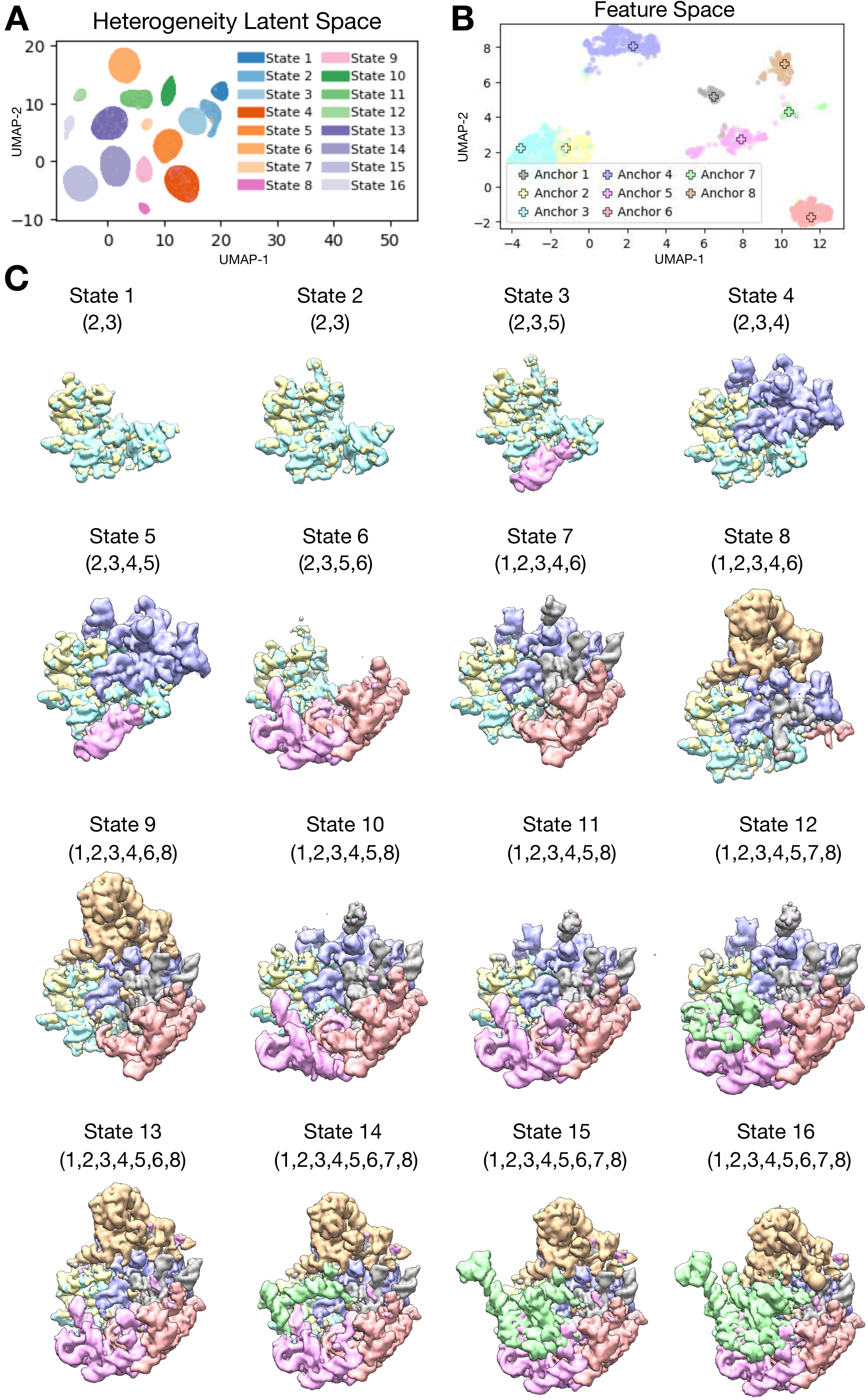}
    \caption{Complete qualitative results on Ribosembly~\cite{jeon2024cryobench}
    \textbf{(A)} Gaussian feature space, $\fSpace$, showing eight major parts identified through clustering.
    \textbf{(B)} Heterogeneity latent space, $\zSpace$, colored coded with the ground-truth compositional state.
    \textbf{(C)} Visualizations of all 3D density maps corresponding to $16$ compositional states, with colors depicting parts (given in parentheses).
    }
    \vspace*{-0.25cm}
    \label{fig:full-Ribosembly}
\end{figure}

While Figure~\ref{fig:Ribosembly} in the main body of the paper shows example compositional states reconstructed by \methodName{}, 
Figure~\ref{fig:full-Ribosembly} shows all 16 states.
% \DavidF{avoid saying this if possible?}
% \Shayan{Just wanted to be frank and I get this might have negative connotation. Instead, we can leave it to the rebuttal in case they noticed, though that's very probable.}
% Unfortunately there was inconsistency between numbers in the parentheses in panel C and anchor in panel B.  
% Fig.~\ref{fig:full-Ribosembly}, shows all compositional states with corrected part numbers.
% \DavidF{State 9 looks wrong ... are 9 and 10 the same?}
% \Shayan{I just fixed the labels. There was redudnacy. But 9 and 10 are not the same. 
% You can see the difference in the part colored pink (anchor 6).
% Actually states 9, 10, 11 are pretty close to each other and in Fig 11 last row we show that 3dva and drgnai fail to differentiate state 11 from 9 and 10.}

\clearpage
\section{Qualitative Comparisons on CryoBench Data}
\vspace*{-0.2cm}

\begin{figure}[h!]
    \centering
    \includegraphics[width=\linewidth]{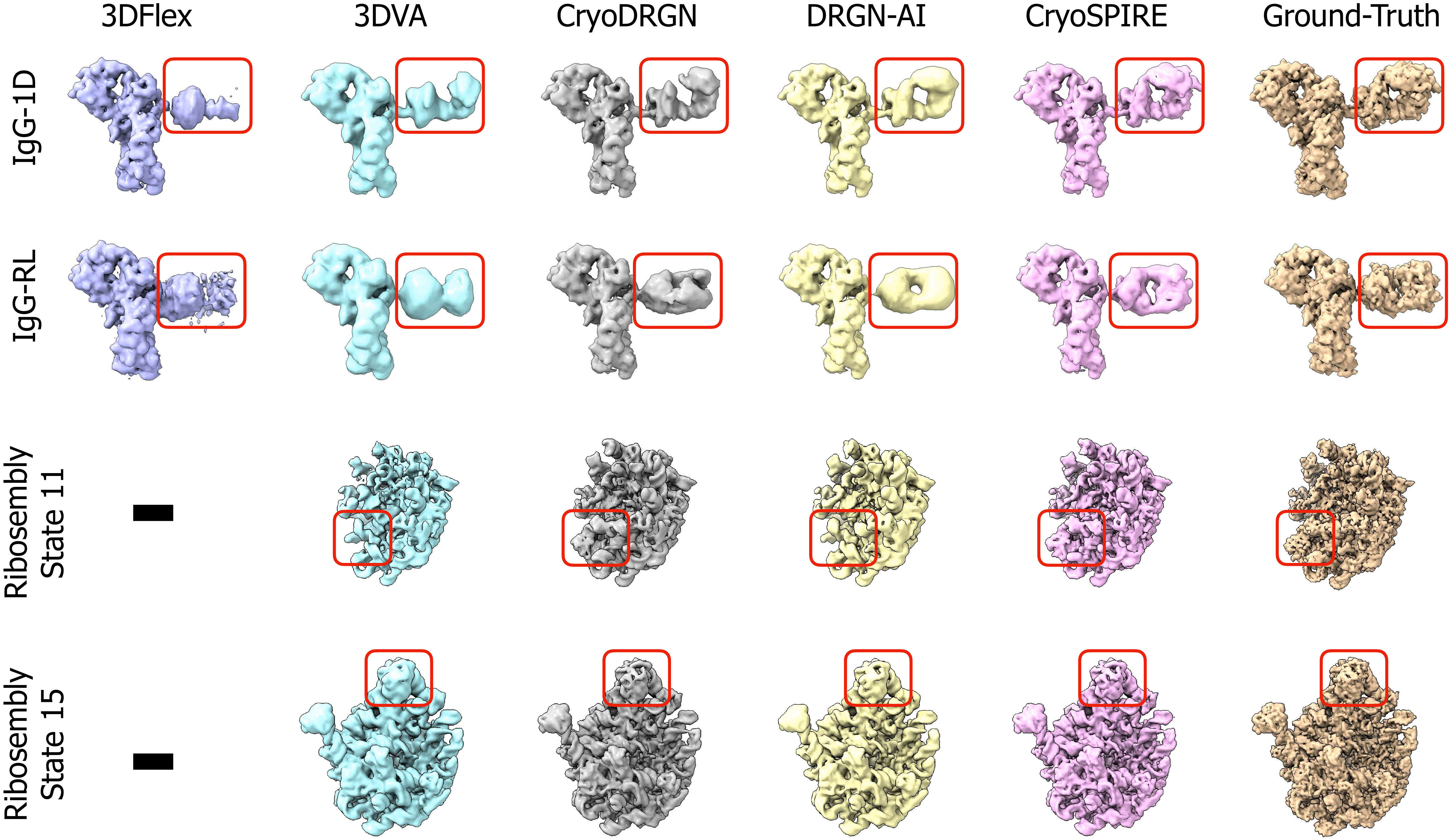}
    \caption{Qualitative comparison of CryoSPIRE with four state-of-the-art methods on CryoBench synthetic datasets~\cite{jeon2024cryobench}.
    Last column corresponds to the ground-truth state.
    In the first two rows, reconstruction of all methods for a sample conformational state is provided for IgG-1D and IgG-RL, demonstrating that our method outperforms others in recovering higher frequency details in the Fab domain (highlighted in red).
    For Ribosembly, we provide reconstructions of two example compositional states (labeled as 11 and 15).
    Since 3DFlex is limited to conformational heterogeneity, it is not evaluated on this dataset.
    Reconstructions of the state 11 by DRGN-AI and 3DVA clearly miss a subunit, while CryoSPIRE is able to capture it.
    Moreover, for state 15, DRGN-AI and 3DVA are overall less detailed while CryoSPIRE appears slightly better than CryoDRGN.
    3D visualizations of the above reconstructions are presented in the webpage.
    }
    \label{fig:all-methods-synthetic}
\end{figure}

In Figures~\ref{fig:IgG-1D},~\ref{fig:IgG-RL} and~\ref{fig:Ribosembly}, we showed qualitative result of \methodName{} on CryoBench synthetic datasets of IgG-1D, IgG-RL and Ribosembly, respectively.
Here, in Figure~\ref{fig:all-methods-synthetic}, we compare \methodName\ with the state-of-the-art methods 3DFlex~\cite{3DFlex-NM-2023}, 3DVA~\cite{3DVA2021}, CryoDRGN~\cite{cryoDRGN2021}, DRGN-AI~\cite{levy2024revealing}, and with ground-truth structural states.
Please find detailed 3D visualizations of the reconstruction in the webpage. 
\vspace*{0.1cm}

\clearpage
\section{Qualitative Comparisons on Experimental datasets}
\label{sec:supp-qual-10076}
\vspace*{-0.2cm}

\begin{figure}[h!]
    \centering
    \includegraphics[width=0.99\linewidth]{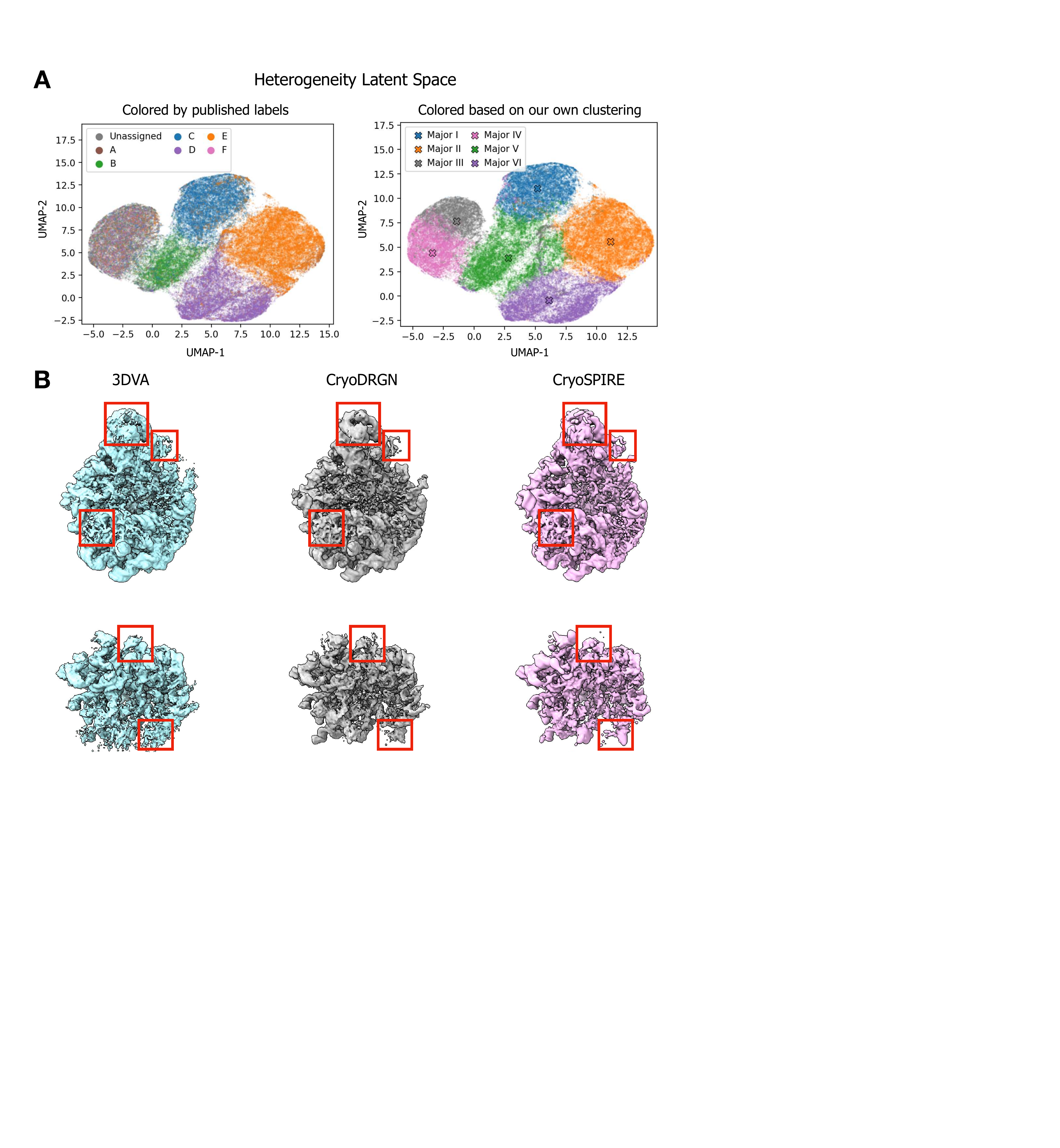}
    \caption{
    More qualitative result on Large Ribosomal Subunit (EMPIAR-10076~\cite{davis2016modular})
    \textbf{(A)} Heterogeneity latent space, with latent points colored based on the published labels~\cite{davis2016modular} (left) and colored based on our own clustering of latent space (right).
    \textbf{(B)} Qualitative comparison of CryoSPIRE with 3DVA~\cite{3DVA2021} and CryoDRGN~\cite{cryoDRGN2021}.
    Two rows shows two of major assembly states.
    We identify some areas with red rectangles that shows main discrepancies between different methods. 
    3D visualizations of the above reconstructions are presented in the webpage.
    }
    \label{fig:more-10076}
\end{figure}

\begin{figure}[h!]
    \centering
    \includegraphics[width=\linewidth]{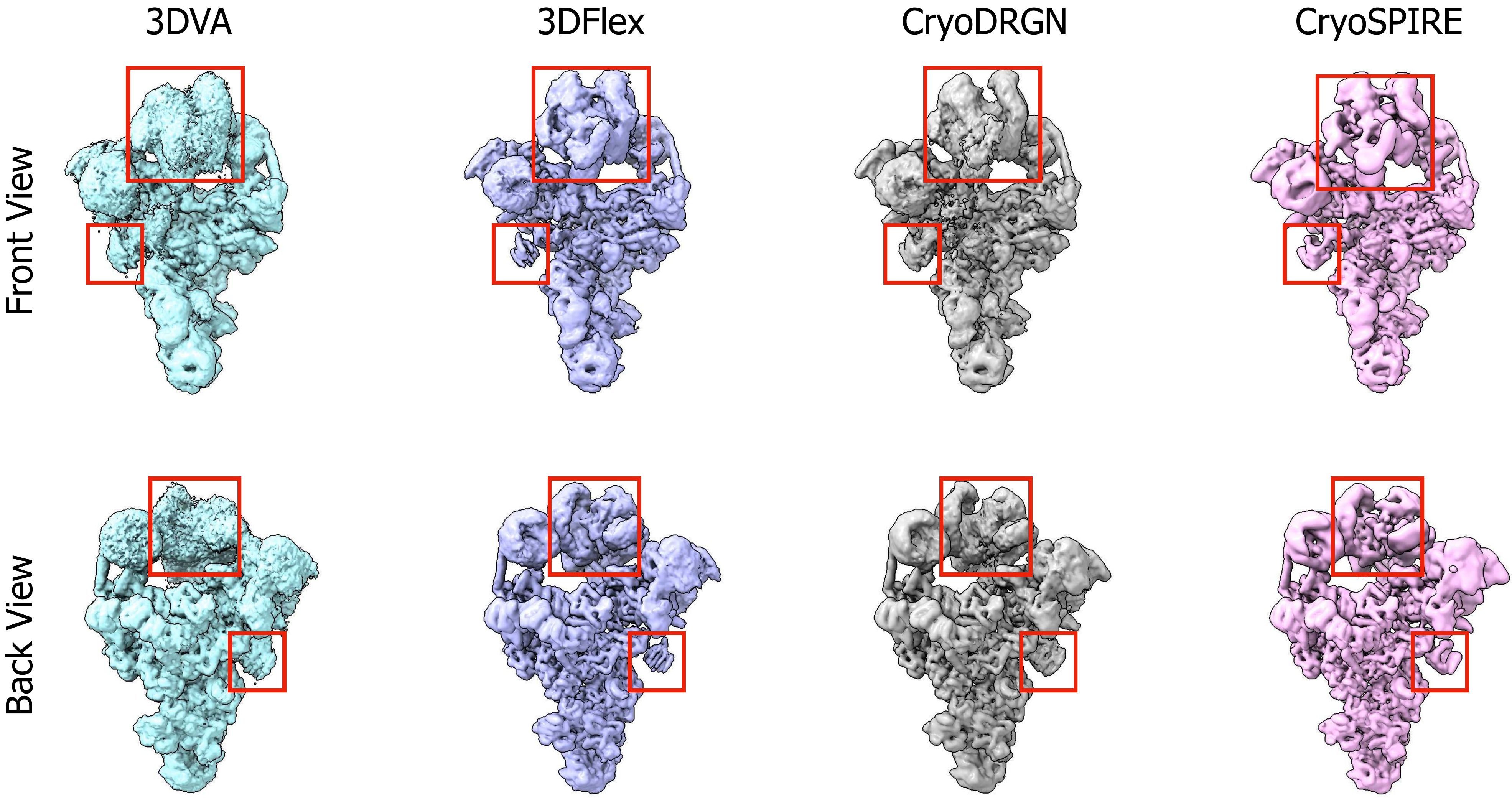}
    \caption{
    Qualitative comparison of CryoSPIRE with 3DVA~\cite{3DVA2021}, 3DFlex~\cite{3DFlex-NM-2023} and CryoDRGN~\cite{cryoDRGN2021} on Pre-Catalytic Spliceosome (EMPIAR-10180~\cite{plaschka2017structure}).
    Two rows show front and back views of the reconstructions, respectively.
    In both views, we mark the SF3b and a peripheral subunit of helicase with red rectangles.
    The reconstructions obtained by 3DVA, 3DFlex and CryoDRGN contain high-frequency noise within the two marked areas, whereas our method is better in resolving corresponding regions.
    3D visualizations of the above reconstructions are presented in the webpage.
    }
    \label{fig:all-methods-10180}
\end{figure}

Figures~\ref{fig:10076-result} and~\ref{fig:10180-result} in the main body of the paper showed qualitative result of \methodName{} on Large Ribosomal Subunit (EMPIAR-10076~\cite{davis2016modular}) and Pre-Catalytic Spliceosome (EMPIAR-10180~\cite{plaschka2017structure}), respectively.
Here, in Figs~\ref{fig:more-10076} and ~\ref{fig:all-methods-10180}, we show comparsions with state-of-the-art methods 3DVA~\cite{3DVA2021} and CryoDRGN~\cite{cryoDRGN2021} and 3DFlex~\cite{3DFlex-NM-2023}.
As these Ribosome data mainly exhibit compositional heterogeneity, 3DFlex is not evaluated on this dataset.
For both 3DVA and 3DFlex, we use CryoSPARC v4.4.0~\cite{cryoSPARC-NM-2017} with default setting.
We use the default $3$ variability components for 3DVA, while for 3DFlex, we run {\em 3DFlex Training} Job, followed by {\em 3DFlex Reconstruction} to obtain a high-resolution canonical structure.
For CryoDRGN, we use the final result provided by the authors.
Please find detailed 3D visualizations of the reconstruction in the webpage.
\vspace*{0.1cm}

\vfill

\vfill
%%%%%%%%%%%%%%%%%%%%%%%%%%%%%%%%%%%%%%%%%%%%%%%%%%%%%%%%%%%%

\end{document}